\documentclass[aos,MSNbibl,nameyear,dvips]{arximspdf}
\usepackage{algorithmic,algorithm}
\usepackage{graphicx}

\doi{10.1214/13-AOS1096} 
\volume{41}
\issue{3}
\pubyear{2013}
\firstpage{1111}
\lastpage{1141}

\makeatletter
\newproclaim{remark}{Remark}
\def\cal{\mathcal}
\newcommand{\eqref}[1]{(\ref{#1})}
\newtheorem{property}{Property}
\newtheorem{lemma}{Lemma}
\newcommand{\bp}{\beta^+}
\newcommand{\bn}{\beta^-}
\newcommand{\gp}{\gamma^+}
\newcommand{\gn}{\gamma^-}
\newcommand{\hbp}{\hat\beta^+}
\newcommand{\hbn}{\hat\beta^-}
\newcommand{\tX}{\widetilde{X}}
\newcommand{\A}{\mathcal{A}}
\renewcommand{\L}{\mathcal{L}}
\newcommand{\tr}{\operatorname{tr}}
\newcommand{\hTheta}{\widehat{\Theta}}
\newcommand{\real}{\mathbb{R}}
\newcommand{\Minimize}{\mathrm{Minimize}}
\newcommand{\argmin}{\operatorname{argmin}}
\newcommand{\N}{\mathcal{N}}
\newcommand{\T}{\mathcal{T}}
\renewcommand{\L}{\mathcal{L}}

\newcommand{\B}{\mathcal{B}}
\newcommand{\hmu}{\hat\mu}
\newcommand{\hnu}{\hat\nu}
\newcommand{\hphi}{\hat\phi}
\newcommand{\ns}{\mathrm{null}}
\renewcommand{\P}{\mathcal{P}}
\newcommand{\Z}{\mathcal{Z}}
\newcommand{\tD}{\widetilde{D}}

\def\implies{\Longrightarrow}

\setattribute{abstract}{width}{290pt}
\makeatother

\begin{document}
\begin{frontmatter}

\title{A lasso for hierarchical interactions}
\runtitle{Hierarchical interactions lasso}

\begin{aug}
\author[A]{\fnms{Jacob} \snm{Bien}\corref{}\thanksref{t1}\ead[label=e1]{jbien@cornell.edu}},
\author[B]{\fnms{Jonathan} \snm{Taylor}\thanksref{t2}\ead[label=e2]{jonathan.taylor@stanford.edu}}
\and
\author[C]{\fnms{Robert} \snm{Tibshirani}\thanksref{t3}\ead[label=e3]{tibs@stanford.edu}}
\thankstext{t1}{Supported in part by the Gerald J. Lieberman Fellowship.}
\thankstext{t2}{Supported by NSF Grant DMS-09-06801
and AFOSR Grant 113039.}
\thankstext{t3}{Supported by NSF Grant
DMS-99-71405 and National Institutes of Health Contract N01-HV-28183.}
\runauthor{J. Bien, J. Taylor and R. Tibshirani}
\affiliation{Cornell University, Stanford University and Stanford University}
\address[A]{J. Bien\\
Department of Biological Statistics\\
\quad and Computational Biology\\
and Department of Statistical Science\\
Cornell University\\
Ithaca, New York 14853\\
USA\\
\printead{e1}}
\address[B]{J. Taylor\\
Department of Statistics\\
Stanford University\\
Stanford, California 94305\\
USA\\
\printead{e2}}
\address[C]{R. Tibshirani\\
Department of Health, Research, \& Policy\\
and Department of Statistics\\
Stanford University\\
Stanford, CA 94305\\
USA\\
\printead{e3}}
\end{aug}

\received{\smonth{5} \syear{2012}}
\revised{\smonth{12} \syear{2012}}

%
\begin{abstract}
We add a set
of convex constraints to the lasso to produce sparse interaction models
that honor the hierarchy
restriction that an interaction only be included in a model if
one or both variables are marginally important.
We give a precise characterization of the effect of this hierarchy
constraint, prove that hierarchy
holds with probability one and derive an unbiased estimate for the
degrees of freedom of our estimator. A bound on this estimate
reveals the amount of fitting ``saved'' by the hierarchy constraint.

We distinguish between \emph{parameter
sparsity}---the number of nonzero coefficients---and \emph{practical
sparsity}---the number of raw variables one must \emph{measure} to
make a new prediction. Hierarchy focuses on
the latter, which is more closely tied to important data
collection concerns such as cost, time and effort.
We develop an algorithm, available in the \texttt{R} package
\texttt{hierNet}, and perform an empirical study of our method.
\end{abstract}

%

\end{frontmatter}

\section{Introduction}\label{sec1}
There are numerous situations in which additive (main effects) models are
insufficient for predicting an outcome of interest.
In medical diagnosis, the co-occurrence of two symptoms may lead a
doctor to be confident that a patient has a certain disease whereas
the presence of either symptom without the other would provide only a
moderate indication of
that disease. This situation corresponds to a positive (i.e.,
synergistic) interaction
between symptom variables. On the other hand, suppose both symptoms
convey redundant information to the doctor about the patient so that
knowing both provides no
more information about the disease status than either one on its
own. This situation is again not additive, but this time there is a
negative interaction between symptoms. Fitting regression models with
interactions is challenging when one has even a moderate number, $p$,
of measured variables, since there
are ${p \choose k}$ interactions of order $k$. For this paper, we
focus on the case of pairwise ($k=2$) interaction models, although the
ideas we develop generalize naturally to higher-order interaction
models.

\subsection{Two-way interaction model}
\label{sec:interactions-model}
We consider a regression model for an outcome variable $Y$
and predictors $X_1,\ldots,X_p$, with pairwise interactions
between these predictors.
In particular our model has the form
%
\begin{equation}
\label{eq:model} Y = \beta_0+\sum_{j}
\beta_jX_j + \frac{1}{2}\sum
_{j\neq k}\Theta _{jk}X_jX_k+
\varepsilon,
\end{equation}
where $\varepsilon\sim N(0, \sigma^2)$. Regardless of whether the
predictors are continuous or discrete, we will refer to the additive
part as the ``main effect'' terms and the quadratic part as the
``interaction'' terms. Our goal is to estimate $\beta\in\real^p$ and
$\Theta\in\real^{p\times p}$, where $\Theta=\Theta^T$ and
$\Theta_{jj}=0$. The factor of one half before the interaction
summation is a consequence of our notational decision to deal with a
symmetric matrix $\Theta$ of interactions rather than a vector of length
$p(p-1)/2$. We take $\Theta_{jj}=0$ throughout this paper because it
simplifies notation, but everything carries over if we remove this
restriction. Indeed, we provide this as an option in the \texttt{hierNet}
(pronounced ``hair net'') package.

We observe a training sample, $(x_1,y_1), \ldots, (x_n,y_n)$, and our
goal is to select a subset of the $p+p(p-1)/2$ main effect and
interaction variables that is predictive of the response, and to
estimate the values for the nonzero parameters of the model.

\subsection{Strong and weak hierarchy}
\label{sec:strong-weak-hier}
It is a well-established practice among statisticians fitting
\eqref{eq:model} to only allow an interaction into the model if the
corresponding main effects are also in the model. Such restrictions
are known under various names, including ``heredity,''
``marginality,'' and being
``hierarchically well-formulated'' [\citet
{Hamada92,Chipman96,Nelder77,Peixoto87}]. There are two types of
restrictions, which we will call \emph{strong} and \emph{weak hierarchy}:
\begin{eqnarray*}
\mbox{\textsc{Strong hierarchy:}}&&\quad \hTheta_{jk}\neq0\quad\implies\quad \hat
\beta_j\neq0\quad\mbox{and}\quad\hat\beta_k\neq0;
\\
\mbox{\textsc{Weak hierarchy:}}&&\quad \hTheta_{jk}\neq0\quad\implies \quad\hat
\beta_j\neq0\quad\mbox{or}\quad\hat\beta_k\neq0.
\end{eqnarray*}
Some statisticians argue that models violating strong hierarchy are
not sensible. For example, according to \citet{McCullagh83},

\begin{quote}
``[T]here is usually no reason to postulate a special position for
the origin, so that the linear terms must be included with the cross-term.''
\end{quote}

To see that violating strong hierarchy amounts to ``postulating a
special position for the origin,'' consider writing an interaction
model as $Y=\beta_0+(\beta_1+\Theta_{12}X_2)X_1+\cdots$. First of
all, we would only take $\beta_0=0$ if we have special reason to
believe that the regression surface must go through the origin.
Likewise, taking $\beta_1=0$ but $\Theta_{12}\neq0$ would only be appropriate
if we actually believe that $X_1$'s effect on $Y$ should only be present
specifically when $X_2$ is nonzero. In most situations, we do not
think that the variable $X_2$ that we measured is any more special than
$aX_2+b$. Yet if our model with $X_2$ violates strong hierarchy, then
our model with $aX_2+b$ (for any $b\neq0$) is strongly hierarchical.
This argument suggests that violations to hierarchy occur in special
situations whereas hierarchy is the default.

Another argument in favor of hierarchy has to do with statistical
power. In the words of \citet{Cox84}:

\begin{quote}
``[L]arge component main effects are more likely to lead to
appreciable interactions than small components. Also, the
interactions corresponding to larger main effects may be in some
sense of more practical importance.''
\end{quote}

In other words, rather than looking at all possible interactions, it
may be useful to focus our search on those interactions that have
large main effects. Indeed, the method we propose in this paper makes
direct use of this principle.

\begin{figure}

\includegraphics{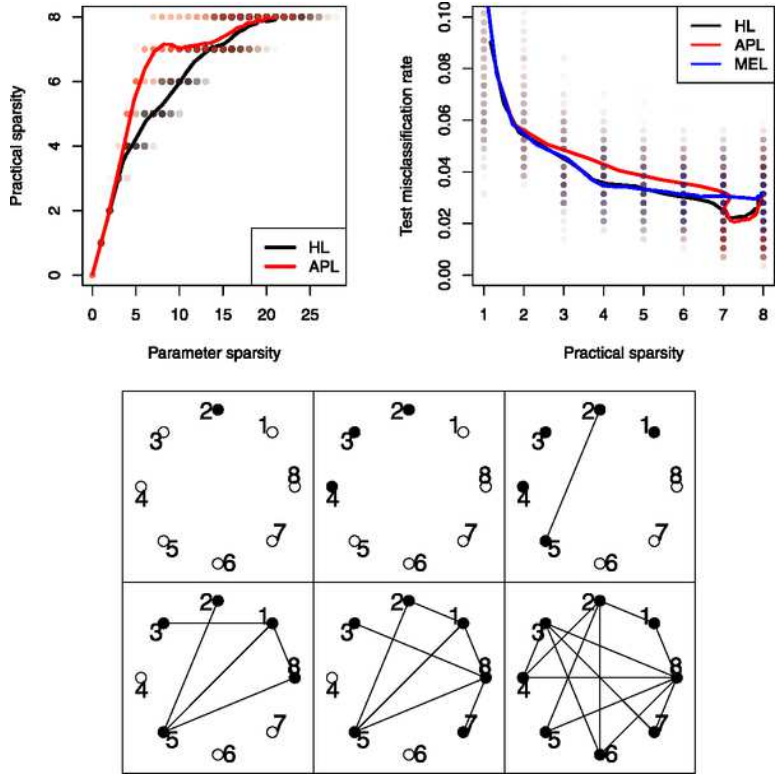}

\caption{Olive oil data: (Top left) Parameter sparsity is the number
of nonzero coefficients while practical sparsity is the number of
\emph{measured} variables in the model. Results from all 100 random
train-test splits are
shown as points; lines show the average performance
over all 100 runs. (Top right) Misclassification error on test
set versus practical sparsity. (Bottom) Wheel plots showing the sparsity
pattern at 6 values of $\lambda$ for the strong hierarchical
lasso. Filled nodes correspond to nonzero main effects, and
edges correspond to nonzero interactions.}
\label{fig:olive}
\end{figure}

As a final argument for hierarchy, it is useful to distinguish between
two notions of sparsity,
which we will call \emph{parameter
sparsity} and \emph{practical sparsity}. Parameter sparsity is what
most statisticians mean by ``sparsity'': the number of nonzero
coefficients in the
model. Practical sparsity is what someone actually collecting data
cares about:
the number of variables one needs to \emph{measure} to make predictions
at a future time. The hierarchy restriction favors models that
``reuse'' measured
variables whereas a nonhierarchical model does not. The top left panel of
Figure~\ref{fig:olive} gives a small example where this difference is
manifest. In fact, a simple calculation shows that this difference
can be quite substantial: we can have a
hierarchical and a nonhierarchical interaction model with the same
parameter sparsity but with the nonhierarchical method having a
practical sparsity of $k(k+1)$ whereas the hierarchical method's
practical sparsity is just $k$.

While taking these arguments to the extreme leads to the use of strong
hierarchy exclusively, we develop the case of weak hierarchy in parallel
throughout this paper. Weak hierarchy, as the name suggests, can be
thought of as a
compromise between strong hierarchy and imposing no such structure and
appears as a principle in certain statistical methods such as
classification and regression trees [\citet{BFOS84}] and multivariate
additive regression splines [\citet{Friedman91}].

\subsection{Sparsity, the lasso and structured sparsity}
\label{sec:sparsity-lasso}

The lasso [\citet{Tibshirani96}] is a method that performs both
model selection and estimation. It penalizes the squared loss of
the data with an $\ell_1$-norm penalty on the parameter vector.
This penalty has the property of producing estimates of the parameter
vector that are sparse (corresponding to model selection). Given a
design matrix $\tX\in\real^{n\times d}$ and response vector
$y\in\real^n$, the lasso is the solution to the convex optimization
problem,
\[
\mathop{\Minimize}_{\beta_0,\phi} \frac{1}{2}\Vert y-\beta_01-\tX\phi
\Vert ^2+\lambda\Vert\phi\Vert_1,
\]
where $1\in\real^n$ is the vector of ones. The penalty parameter,
$\lambda\ge0$, controls the relative importance of
fitting to the training data (sum-of-squares term) and of sparsity
($\ell_1$
penalty term). A natural extension of the lasso to our interaction model~\eqref{eq:model} would be to take $\phi^T=[\beta^T,\operatorname
{vec}(\Theta
)^T]$ and
$\tX=(X:Z/2)$, where the columns of $Z\in\real^{n\times p(p-1)}$
correspond to elementwise
products of the columns of $X$. We will refer to this method as the
\emph{all-pairs lasso} since it is simply the lasso applied to a data
matrix which includes all
pairs of interactions (as well as all main effects). It is common
with the lasso to standardize the predictors so that they are on the
same scale. In this paper, we standardize $X$ so that its columns have
mean 0 and standard deviation~1; we then form $Z$ from these
standardized predictors and, finally, center the resulting columns of
$Z$. By centering $y$ and $\tX$, we may take $\hat\beta_0=0$.

The lasso's $\ell_1$ penalty is neutral to the pattern of sparsity,
allowing any sparsity pattern to emerge. The notions of strong and
weak hierarchy introduced in Section~\ref{sec:strong-weak-hier}
represent situations in which we want to
exclude certain sparsity patterns. There has been a growing
literature focusing on methods that produce \emph{structured sparsity}
[\citet{Yuan06},
\citet{Zhao09},
\citet{Jenatton10},
\citet{Jenatton11},
\citet{Bach11},
\citet{Bach12}]. These
methods make use of
the \emph{group lasso} penalty (and generalizations thereof)
which, given a predetermined grouping of the parameters, induces
entire groups of parameters to be set to zero [\citet{Yuan06}].
Given a set of groups of variables, $\mathcal G$, these methods
generalize the $\ell_1$ penalty by
\[
\sum_{G\in\mathcal G}d_G\Vert\phi_G
\Vert_{\gamma_G},
\]
where $\gamma_G>1$, $\phi_G$ is $\phi$ projected onto the
coordinates in $G$, and $d_G$ is a nonnegative weight. Hierarchical
structured sparsity is obtained by
choosing $\mathcal G$ to have nested groups. For example,
\citet{Zhao09} consider the penalty
\[
\sum_{j\neq k} \bigl\{|\Theta_{jk}|+\bigl\Vert(
\beta_j,\beta_k,\Theta _{jk})
\bigr\Vert_{\gamma_{jk}} \bigr\}.
\]
Likewise, the framework of \citet{Bach12} if specialized to this paper's
focus would lead to a penalty of the form
%
\begin{equation}
\label{eq:bach} \Vert\Theta\Vert_1+\sum
_{j}d_j\bigl\Vert(\Theta_j,
\beta_j)\bigr\Vert_q
\end{equation}
for some $q>1$ and $d_j>0$. In fact, \citet{Radchenko10} suggest a
penalty for generalized additive models with interactions that reduces
to \eqref{eq:bach} in the linear model case, with $q=2$ and $d_j$
independent of $j$.

\subsection{This paper}
\label{sec:this-paper}
Here, we propose a lasso-like procedure that produces sparse
estimates of $\beta$ and $\Theta$ while satisfying the strong or weak
hierarchy
constraint. In contrast to much
of the structured sparsity literature
which is based on group lasso penalties,
our approach, presented in Section~\ref{sec:our-proposed-method},
involves adding a set of convex constraints to the
lasso. Although we find this form of constraint more naturally
interpretable, we show (Remark~\ref{re3}) that this problem can be equivalently
expressed in a form that
relates it to penalties from the structured sparsity literature such as~\eqref{eq:bach}.

A key advantage of our specific choice of penalty structure is that it
admits a simple
interpretation of the effect of the hierarchy demand. Unlike other
hierarchical sparsity methods, which do not pay much attention to the
particular choice of norms (as long as $\gamma_G>1$), our formulation
is carefully tailored
to allow it to be related directly back to the lasso,
permitting one to understand specifically how hierarchy alters the
solution (Section~\ref{sec:effect-constraint}). This feature of our
estimator gives it a transparency that exposes the effects (both
positive and negative!) of the hierarchy constraint.
Furthermore, our characterization suggests that the demand for
hierarchy is---analogous to the demand for sparsity---a form of
``regularization.'' We develop an unbiased estimator of
the degrees of freedom of our method (Section~\ref{sec:degrees-freedom}) and an interpretable upper bound on this
quantity, which also points
to hierarchy as regularization. In particular, we show that we do not
``spend'' in degrees of freedom for main effects that are forced into
the model by the hierarchy constraint.

Another difference from much of the structured
sparsity literature, which aims to develop a broad treatment of
structured and hierarchical sparsity methods, is that our focus is
narrowed to the problem of interaction models. Our restricted scope
allows us to address specifically the performance of such a tool to
this important problem. In Section
\ref{sec:related-work}, we review previous work on the problem of
hierarchical interaction model fitting and selection. These
methods fall into three categories: Multi-step procedures, which are defined
by an algorithm
[\citet{Peixoto87},
\citet{Friedman91},
\citet{Turlach04},
\citet{Nardi07},
\citet{Bickel08},
\citet{Park08},
\citet{Wu10}];
Bayesian approaches, which specify the hierarchy requirement through a
prior [\citet{Chipman96}]; and, most related to this paper's proposal,
regularized regression methods, which are defined by an optimization
problem [\citet{Yuan09},
\citet{Zhao09},
\citet{Choi10},
\citet{Jenatton10},
\citet{Radchenko10}].
In Section~\ref{sec:empirical-study}, we study via simulation the statistical
implications of imposing hierarchy on an interactions-based estimator
under various
scenarios (in both the lasso and stepwise frameworks). In Section~\ref{sec:algorithmics} we present an efficient algorithm for computing
our estimator. Real data examples are used to
illustrate a distinction we draw between ``parameter sparsity'' and
``practical sparsity'' and to discuss hierarchy's role in promoting the latter.

\section{Our proposed method}
\label{sec:our-proposed-method}

In Section~\ref{sec:sparsity-lasso}, we introduced the \emph{all-pairs
lasso}, which can be written as
%
\begin{equation}
\label{eq:APL} \mathop{\Minimize}_{\beta_0\in\real, \beta\in\real^p, \Theta\in\real
^{p\times
p}}  q(\beta_0,\beta,\Theta) +
\lambda\Vert\beta\Vert_1 + \frac{\lambda}{2}\Vert\Theta
\Vert_1 \qquad\mbox{s.t.}\qquad \Theta= \Theta^T,
\end{equation}
where $\Vert\Theta\Vert_1=\sum_{j\neq k}|\Theta_{jk}|$ and
$q(\beta
_0,\beta,\Theta)$ is the loss function, typically\break $\frac{1}{2}\sum_{i=1}^n(y_i -\beta_0-
x_i^T\beta- \frac{1}{2}x_i^T\Theta x_i)^2=\frac{1}{2}\Vert
y-\beta_01-X\beta-Z\operatorname{vec}(\Theta)/2\Vert^2$, but may also include
a ridge penalty on the coefficients as discussed later or may be
substituted for the binomial negative log-likelihood. The one-half
factors in front of terms involving $\Theta$ are merely a consequence of
the notational choice to represent $\Theta$ as a symmetric
matrix (with $\Theta_{jj}=0$ for $j=1,\ldots,p$). In this paper, we
propose a modification of
the all-pairs lasso that produces models that are guaranteed to be
hierarchical.

As motivation for our proposal, consider building hierarchy into the
optimization problem as a constraint,
%
\begin{eqnarray}
\label{eq:nonconvex}&& \mathop{\Minimize}_{\beta_0\in\real, \beta\in\real^p, \Theta\in\real
^{p\times
p}}  q(\beta_0,\beta,
\Theta) + \lambda\Vert\beta\Vert_1 +\frac{\lambda}{2}\Vert\Theta
\Vert_1
\nonumber
\\[-8pt]
\\[-8pt]
\nonumber
&&\qquad\mbox{s.t.}\qquad \Theta= \Theta^T, \Vert\Theta_j
\Vert_1\le|\beta_j| \qquad\mbox{for }j=1,\ldots,p,
\end{eqnarray}
where $\Theta_j$ denotes the $j$th row (and column, by symmetry) of
$\Theta$. Notice that if $\hTheta_{jk}\neq0$, then $\Vert\hTheta
_j\Vert
_1>0$ and
$\Vert\hTheta_k\Vert_1>0$ and thus $\hat\beta_j\neq0$ and
$\hat\beta_k\neq0$. While the added constraints enforce strong
hierarchy, they are not convex, which makes~\eqref{eq:nonconvex}
undesirable as a method. In this
paper, we propose a straightforward
convex relaxation of \eqref{eq:nonconvex}, which we call the
\emph{strong hierarchical lasso},
%
\begin{eqnarray}
\label{eq:hiernet} &&\mathop{\operatorname{Minimize}\limits_{\beta_0\in\real, \beta^\pm\in\real
^p,}}_{\Theta\in \real^{p\times p}} q\bigl(
\beta_0,\bp-\bn,\Theta\bigr) + \lambda 1^T\bigl(\bp+\bn\bigr) +
\frac{\lambda}{2}\Vert\Theta\Vert_1
\nonumber
\\[-8pt]
\\[-8pt]
\nonumber
&&\qquad\mbox{s.t.} \qquad\Theta= \Theta^T,  %
\left.\begin{aligned}
\Vert\Theta_j\Vert_1&\le\bp_j+
\bn_j
\\
\bp_j&\ge0, \bn_j\ge0
\end{aligned} %
\right\}\mbox{ for }j=1,\ldots,p,
\nonumber
\end{eqnarray}
where we have replaced the optimization variable $\beta\in\real^p$ by
two vectors $\bp,\bn\in\real^p$.
After solving the above problem, our fitted model is of the form
$\hat f(x)=\hat\beta_0+x^T(\hbp-\hbn)+x^T\hTheta x/2$. While we might
informally think of $\bp$ and $\bn$ as positive and negative parts of
a vector $\beta=\bp-\bn$, that is, that $\beta^\pm=\max\{\pm
\beta,0\}$,
this is not actually the case since at a solution we can have both
$\hbp
_j>0$ and
$\hbn_j>0$. Indeed, if we were to add the constraints $\bp_j\bn_j=0$
for $j=1,\ldots,p$ to \eqref{eq:hiernet}, then these would be positive
and negative parts and so $\bp_j+\bn_j=|\beta_j|$, giving us
precisely problem
\eqref{eq:nonconvex}. This observation establishes that
\eqref{eq:hiernet} is a convex relaxation of~\eqref{eq:nonconvex}.

The hierarchy constraints can be seen as an embedding into our method
of David Cox's
``principle'' that ``large component main effects are more likely to
lead to appreciable interactions than small components.'' The
constraint
\[
\Vert\Theta_j\Vert_1\le\bp_j+
\bn_j
\]
budgets the total amount of interactions involving
variable $X_j$ according to the relative importance of $X_j$ as a main
effect. One additional advantage of the convex relaxation\vadjust{\goodbreak} is that the
constraint is less restrictive. If the best fitting model would have
$\Vert\Theta_{j}\Vert_1$ large but $|\beta_j|$ only moderate, this
can be accommodated by making $\bp_j$ and $\bn_j$ both large.

\begin{remark}\label{rem1}
Another possibility for the
hierarchy constraint
that we have considered is $|\Theta_{jk}|\le\bp_j+\bn_j$; however, we
have found that this can lead to an overabundance of interactions
relative to main effects.
\end{remark}

\begin{remark}\label{rem2}
It is desirable to include in the loss
function $q$ an \emph{elastic net} term,
$(\varepsilon/2) (\Vert\Theta\Vert_F^2+\Vert\bp\Vert^2+\Vert
\bn\Vert
^2 )$,
to ensure uniqueness of the solution [\citet{Zou05}]. We think of
$\varepsilon>0$ as a fixed
tiny fraction of $\lambda$, such as $\varepsilon=10^{-8}\lambda$, rather
than as an additional
tuning parameter. Such a modification does not complicate the
algorithm, but simplifies the study of the estimator. In all numerical
examples and
in the \texttt{hierNet} package, we use this elastic net modification.
\end{remark}

\begin{remark}\label{re3}
We prove in Section~2 of
the supplementary materials [\citet{Bien13supp}] that \eqref
{eq:hiernet} may equivalently be
written as
%
\begin{equation}
\label{eq:will} \mathop{\operatorname{Minimize}\limits_{\beta_0\in\real, \beta\in\real^p,}}\limits_{\Theta
\in
\real^{p\times p},\Theta=\Theta^T} q(
\beta_0,\beta,\Theta) + \lambda\sum_j
\max\bigl\{|\beta_j|,\Vert\Theta_j\Vert_1\bigr\}+
\frac
{\lambda
}{2}\Vert\Theta\Vert_1.
\end{equation}
This reparameterization of the problem shows its similarities to the
group lasso based methods. In place of the more standard penalty $\Vert
(\Theta_j,\beta_j)\Vert_q$
of \eqref{eq:bach}, we use
$\max\{\Vert\Theta_j\Vert_1,|\beta_j|\}$. In Section
\ref{sec:effect-constraint}, we show that this unusual choice of
penalty admits a particularly simple interpretation for the effect of
imposing hierarchy.
\end{remark}

In Section~\ref{sec:strong-weak-hier}, we also introduced the notion
of weak hierarchy. By simply removing the symmetry constraint on
$\Theta$, we get what we call the \emph{weak hierarchical lasso},
%
\begin{eqnarray}
\label{eq:weak-hiernet}&& \mathop{\Minimize}_{\beta_0\in\real, \beta^\pm\in\real^p, \Theta\in
\real
^{p\times p}} q\bigl(\beta_0,\bp-\bn,
\Theta\bigr) + \lambda1^T\bigl(\bp +\bn\bigr) +\frac{\lambda}{2}\Vert\Theta
\Vert_1
\nonumber
\\[-8pt]
\\[-8pt]
\nonumber
&&\qquad\mbox{s.t.}\qquad
\left.\begin{aligned} \Vert\Theta_j
\Vert_1&\le\bp_j+\bn_j
\\
\bp_j&\ge0, \bn_j\ge0
\\
\end{aligned} %
\right\}\mbox{ for }j=1,\ldots,p.
\end{eqnarray}
Even though at a solution to this problem, $\hTheta$ is not symmetric,
we should think of the interaction coefficient as
$(\hTheta_{jk}+\hTheta_{kj})/2$ since this is what multiplies
the interaction term $x_{ij}x_{ik}$ when computing $\hat f(x_i)$.

\begin{remark}\label{rem4}
We can build further on the
connection between
\eqref{eq:bach} and \eqref{eq:hiernet} discussed in Remark~\ref{re3}. Our
removal of the
symmetry constraint in \eqref{eq:weak-hiernet} is analogous to the
technique of duplicating columns of the design matrix used in the
overlap group lasso [\citet{Obozinski11}].\vadjust{\goodbreak}
\end{remark}

A favorable property that distinguishes our method from previous
approaches discussed in Section~\ref{sec:related-work} is the
relative transparency of the role that the hierarchy constraint plays
in our estimator. This aspect is developed in Section \ref
{sec:effect-constraint}.

Although our primary focus in this paper is on the Gaussian setting of
\eqref{eq:model}, our proposal extends straightforwardly to other
situations, such as the logistic regression setting in which the
response is binary. In this case, we simply have
$q(\beta_0,\beta,\Theta)$ be the appropriate negative log-likelihood,
$-\sum_{i=1}^ny_i\log p_i + (1-y_i)\log(1-p_i)$, where
$p_i=[1+\exp(-\beta_0-x_i^T\beta-\frac{1}{2}x_i^T\Theta x_i)]^{-1}$.
In Section~3 of the supplementary materials [\citet
{Bien13supp}], we show
that solving this problem requires only a minor modification to our
primary algorithm. It should also be noted that our estimator (and the
algorithms developed to compute it) is designed for both the $p<n$
and $p\ge n$ setting.

As a preliminary example, consider predicting whether a sample of olive
oil comes from Southern Apulia
based on measurements of the concentration of $p=8$ fatty acids
[\citet{Forina83}]. The dataset consists of $n=572$ samples, and we
average our results over 100 random equal-sized train-test splits. We
compare three methods: (a) a standard lasso with main effects only
(MEL), (b)
the all-pairs lasso (APL), and (c) the strong hierarchical lasso
(HL).

The top left panel of Figure~\ref{fig:olive} shows an interesting
difference between HL and APL. We see that, on average, at a parameter
sparsity level of five, the HL
model uses four of the measured variables whereas APL uses six. Using
the hierarchical model to classify a future olive oil, we only need to
measure four rather than six of the fatty
acids.

The top right panel of Figure~\ref{fig:olive} shows the predictive
performance (versus the practical sparsity) of the three methods. It
appears that HL enjoys the ``best of both worlds,'' matching
the good performance of MEL for low
practical sparsity levels (since it tends to pick out the main effects
first) and the good performance of APL at high
practical sparsity levels (since it can incorporate predictive
interactions). Finally, the bottom panel of
the figure provides a visual display of a sequence of HL's solutions
(by varying $\lambda$). Nonzero main effects are shown as filled nodes,
and edges
indicate nonzero interactions. Since all edges are incident to filled
nodes, we see that strong hierarchy holds.

In the next section, we present several properties of our estimator
that shed light on the effect of adding the convex hierarchy constraint to
the lasso. Among these properties is an unbiased estimate of
the degrees of freedom of our estimator. We view this
degrees-of-freedom result
as valuable primarily for the sake of understanding the effect of
hierarchy. While such an estimate could be used for parameter
selection, we prefer cross validation to select
$\lambda$ since this is more directly tied to the goal of prediction.

\section{Properties}
\label{sec:properties}

\subsection{Effect of the constraint}
\label{sec:effect-constraint}
A key advantage of formulating an estimator as a solution to a
convex problem is that it can be completely characterized by a set of
optimality conditions, known as the Karush--Kuhn--Tucker (KKT)
conditions. These conditions are useful for understanding the effect
that the hierarchy constraint in \eqref{eq:hiernet} and \eqref
{eq:weak-hiernet} has on our solutions. In this section, we will
study the simplest case, taking $q(\beta_0,\beta,\Theta)$ to be the
quadratic loss function with no elastic net
penalty. We let
\begin{eqnarray*}
r^{(-j)}&=&y-\hat y+x_j\hat\beta_j,
\\
r^{(-jk)}&=&y-\hat y+(x_j*x_k) (
\hTheta_{jk}+\hTheta_{kj})/2
\end{eqnarray*}
denote partial residuals (where $*$ denotes elementwise
multiplication, $\hat y$ the vector of fitted values and $x_j$ the
$j$th predictor), and we
assume that $\Vert x_j\Vert_2=1$. For linear regression,
the KKT conditions are known as the normal equations and can be
written as
\[
\hat\beta_j=x_j^Tr^{(-j)},\qquad
\hTheta_{jk}=\frac{(x_j*x_k)^Tr^{(-jk)}}{\Vert x_j*x_k\Vert^2}.
\]
The all-pairs lasso solution satisfies
%
\begin{equation}
\label{eq:lasso} \hat\beta_j= \mathcal S\bigl(x_j^Tr^{(-j)},
\lambda\bigr),\qquad \hTheta_{jk}= \frac{\mathcal S[(x_j*x_k)^Tr^{(-jk)}, \lambda]}{\Vert
x_j*x_k\Vert^2},
\end{equation}
where $\mathcal S$ denotes the soft-thresholding operator defined by
$\mathcal S (c, \lambda)=\break\mathrm{sign}(c)(|c|-\lambda)_+$. Written
this way, we see that the lasso is similar to linear regression, but
all coefficients are shrunken toward 0, with some coefficients (those for
which $|x_j^Tr^{(-j)}|\!\le\!\lambda$) set to zero. It is instructive to
examine the corresponding statements for the strong and weak
hierarchical lasso methods.\looseness=-1

\begin{property}\label{prop:constraint}
The coefficients of the strong and weak hierarchical lassos with
$\lambda>0$ and taking $q(\beta_0,\beta,\Theta)$ to be the quadratic
loss (with no elastic
net penalty) satisfy:
\begin{itemize}
\item\textsc{Strong:}
\begin{eqnarray*}
\hbp_j-\hbn_j&=& \mathcal S\bigl(x_j^Tr^{(-j)},
\lambda-\hat\alpha_j\bigr),
\\
\hTheta_{jk}&=& \frac{\mathcal S[(x_j*x_k)^Tr^{(-jk)}, \lambda+\hat
\alpha_j+\hat\alpha_k]}{\Vert x_j*x_k\Vert^2};
\end{eqnarray*}
\item\textsc{Weak:}
\begin{eqnarray*}
\hbp_j-\hbn_j&= &\mathcal S\bigl(x_j^Tr^{(-j)},
\lambda-\tilde\alpha_j\bigr),
\\
\frac{\hTheta_{jk}+\hTheta_{kj}}{2}&=&\frac{\mathcal
S[(x_j*x_k)^Tr^{(-jk)}, \lambda+2\min\{\tilde\alpha_j,\tilde\alpha
_k\}
]}{\Vert x_j*x_k\Vert^2}
\end{eqnarray*}
\end{itemize}
for some $\hat\alpha_j\ge0$, $j=1,\ldots,p$ with
$\hat\alpha_j=0$ when $\Vert\hTheta_j\Vert_1<\hbp_j+\hbn_j$ (and
likewise for $\tilde\alpha_j$).
\end{property}
\begin{pf}
See Section~1 of the supplementary
materials [\citet{Bien13supp}].
\end{pf}
The $\hat\alpha_j,\tilde\alpha_j$ appearing in the above two properties
are optimal dual variables corresponding
to the $j$th hierarchy constraint for the strong and weak hierarchical
lasso problems, respectively. When $\Vert\hTheta_j\Vert_1 <
\hbp_j+\hbn_j$, we have $\hat\alpha_j=0$ (or $\tilde\alpha_j=0$) by
complementary slackness. Comparing these expressions
to those of the all-pairs lasso gives
insight into the effect of the constraint. Property \ref
{prop:constraint} reveals
that the overall form of the
all-pairs lasso and hierarchical lasso methods is identical. The
difference is that the hierarchy constraint leads to a
reduction in the shrinkage of certain main effects and an
increase in the shrinkage of certain interactions. In particular,
we see that when the hierarchy constraints are loose at the solution,
that is, $\Vert\hTheta_j\Vert_1<\hbp_j+\hbn_j$, the weak
hierarchical lasso's
optimality conditions become identical to the all-pairs
lasso (since $\tilde\alpha_j=0$) for all coefficients involving
$x_j$. For the strong hierarchical lasso, when both the $j$th and
$k$th constraints are loose, the optimality conditions match those of
the all-pairs lasso for the coefficients of $x_j$, $x_k$ and
$x_j*x_k$. The methods differ when constraints are active,
that is, when $\Vert\hTheta_j\Vert_1=\hbp_j+\hbn_j$, which allows
$\hat\alpha_j$ (or $\tilde\alpha_j$) to be nonzero. Intuitively, this
case corresponds to
the situation in which hierarchy would not have held ``naturally''
(i.e., without
the constraint), and the corresponding dual variable plays the role of
reducing $\hTheta_j$ in $\ell_1$-norm and increasing $\hbp_j+\hbn_j$
until the constraint is satisfied. The way in which the weak and strong
hierarchical lasso methods perform this shrinkage is different, but
both are identical to the
all-pairs lasso when all constraints are loose.

\subsection{Hierarchy guarantee}
\label{sec:hierarchy-guarantee}
In Section~\ref{sec:our-proposed-method}, we showed that adding the
constraint $\Vert\Theta_j\Vert_1\le|\beta_j|$ would guarantee that
hierarchy holds. However, we have not yet shown that the same is true of
the convex relaxation's constraint,
$\Vert\Theta_j\Vert_1\le\bp_j+\bn_j$. In particular, while
$\hTheta_{jk}\neq0\implies\hbp_j+\hbn_j\neq0$, we could still have
$\hbp_j-\hbn_j=0$. This would correspond to a model in which
$X_jX_k$ is used in the model, but $X_j$ is not. Intuitively, we
would expect that if $\hbp_j>0$, then $\hbp_j=\hbn_j$ is analogous to
getting an exact zero in linear regression (i.e., a zero probability
event). In this section, we establish that this is in fact the case.

In particular, we study \eqref{eq:hiernet} and
\eqref{eq:weak-hiernet} where $q(\beta_0,\beta,\Theta)$ includes an
elastic net term. The
importance of this modification is that it ensures uniqueness, simplifying
the analysis. As noted in Remark~\ref{rem2}, we think of $\varepsilon$ as a
small, fixed proportion of $\lambda$ rather than as a separate tuning
parameter.
%
\begin{property}\label{prop:strong-hier}
Suppose $y$ is absolutely continuous with respect to the Lebesgue
measure on $\real^n$.
If $(\hbp,\hbn,\hTheta)$ solves \eqref{eq:hiernet}, where
$q(\beta_0,\beta,\Theta)$ is the quadratic loss with an
$\varepsilon>0$ ridge penalty, then strong hierarchy holds with
probability 1, that is,
\[
\hTheta_{jk}\neq0\quad\implies\quad\hbp_j-\hbn_j\neq0
\quad\mbox{and}\quad \hbp _k-\hbn _k\neq0.
\]
\end{property}
\begin{pf}
See Appendix~\ref{app:proof-hier}.
\end{pf}
To understand how dropping the symmetry constraint leads to the ``or''
statement required of weak hierarchy,
note that $X_jX_k$ is in the weak hierarchical lasso model if and only if
$\hTheta_{jk}+\hTheta_{kj}\neq0$. This holds only if
$\hTheta_{jk}\neq0$ or $\hTheta_{kj}\neq0$.

\begin{property}\label{prop:weak-hier}
Suppose $y$ is absolutely continuous with respect to the Lebesgue
measure on $\real^n$.
If $(\hbp,\hbn,\hTheta)$ solves \eqref{eq:weak-hiernet}, where
$q(\beta_0,\beta,\Theta)$ is the quadratic loss with an
$\varepsilon>0$ ridge penalty, then weak hierarchy holds with probability
1, that is,
\[
\frac{\hTheta_{jk}+\hTheta_{kj}}{2}\neq0\quad\implies\quad\hbp_j-\hbn _j\neq0
\quad\mbox{or}\quad \hbp_k-\hbn_k\neq0.
\]
\end{property}
\begin{pf}
See Appendix~\ref{app:proof-hier}.
\end{pf}

\subsection{Degrees of freedom}
\label{sec:degrees-freedom}
In classical statistics, the degrees of freedom of a procedure refer to
the dimension of
the space over which its fitted values can vary. It is useful in that
it provides a measure of how much ``fitting'' the
procedure is doing. This
notion can be generalized to adaptive procedures such as the lasso
[\citet{Stein81},
\citet{Ef86},
\citet{LARS},
\citet{Zou07}]. See (Ryan) \citet{RyanTibs12} for a
thorough discussion. If given data $y\in\real^n$, a procedure $h$
produces fitted values $\hat y=h(y)\in\real^n$, the degrees of freedom
of the procedure
$h$ is defined to be
%
\begin{equation}
\label{eq:df-def} df(h)=\frac{1}{\sigma^2}\sum_{i=1}^n
\operatorname{cov}(y_i,\hat y_i).
\end{equation}

\begin{property}\label{prop:df}
Suppose $y\sim N(\mu,\sigma^2I_n)$. An unbiased estimate of the
degrees of freedom of the strong hierarchical lasso, with quadratic
loss and no ridge penalty, is given by
\[
\widehat{df}_{\lambda}=\mathrm{rank}(\tX P),
\]
where $\tX=(X:-X:Z/2:-Z/2)$ with $Z$ containing the
interactions, and $P$ is a projection matrix which depends on the sign
pattern of $(\hbp,\hbn,\hTheta)$ and on the set of hierarchy constraints
that are tight.
\end{property}

\begin{pf}
See Appendix~\ref{app:df}.
\end{pf}

Figure~\ref{fig:df} provides a numerical evaluation of how well
$\widehat{{df}}_{\lambda}$ estimates
${df}_{\lambda}$. We fix $X\in\real^{n\times
p}, \beta\in\real^p$ and $\Theta\in\real^{p\times p}$, and we generate
$B=10\mbox{,}000$ Monte Carlo
replicates $y^{(1)},\ldots, y^{(B)}\in\real^n$. For each replicate, we
fit the strong
hierarchical lasso along a grid of $\lambda$ values to get
$(\hat\beta_\lambda^{+(b)},\hat\beta_\lambda^{-(b)},\hTheta
_\lambda^{(b)})$
and $\hat y^{(b)}_\lambda\in\real^n$. From these values, we compute
Monte Carlo estimates of
${df}_{\lambda}$ from the definition in
\eqref{eq:df-def} and of $E[\widehat{{df}}_{\lambda}]$.

\begin{figure}

\includegraphics{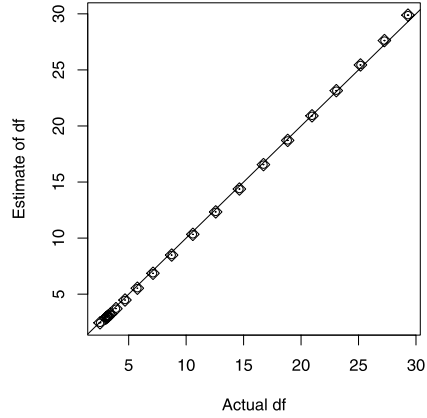}

\caption{Numerical evaluation of how well
$\widehat{{df}}_{\lambda}$ estimates
${df}_{\lambda}$. Monte Carlo estimates of
$E[\widehat{{df}}_{\lambda}]$ ($y$-axis) versus Monte
Carlo estimates of ${df}_{\lambda}$ ($x$-axis) for
a sequence of $\lambda$ values (circular) are shown. One-standard-error bars are
drawn and are hardly visible. Our bound on the unbiased
estimate is plotted with diamonds.}
\label{fig:df}
\end{figure}

While $\widehat{df}_{\lambda}$ can be calculated from the
data and is therefore useful as an unbiased way of calibrating the
amount of fitting the strong hierarchical lasso is doing,\vspace*{1pt} this
expression is difficult to interpret. However, it
turns out that we can \emph{bound} $\widehat{df}_{\lambda}$
by a quantity that does make more sense:
%
\begin{property}\label{prop:df-bound}
Let $\T=\{j\dvtx \Vert\hTheta_j\Vert_1=\hbp_j+\hbn_j\}$,
$\A_\beta=\{j\dvtx \hbp_j-\hbn_j\neq0\},\break \A_\pm=\{j\dvtx \hat\beta^\pm
_j>0\}$
and $\A_\Theta=\{jk\dvtx \hTheta_{jk}\neq0,j<k\}$. Then,
\[
\widehat{df}_{\lambda}\le |\A_\beta|+|\A_\Theta|-\bigl|\T
\cap(\A_+\Delta\A_-)\bigr|
\]
holds almost surely, where $\A_+\Delta\A_-=(\A_+\setminus\A
_-)\cup(\A
_-\setminus\A_+)$.
\end{property}
\begin{pf}
See Appendix~\ref{app:df}.
\end{pf}
By contrast, for the all-pairs lasso in the case that $p+{p\choose2}<n$
and the design matrix is full rank, we have
$df_{\lambda}(\mathrm{APL})=E[|\A_\beta|+|\A_\Theta|]$ [\citet{Zou07}].
In other
words, the strong hierarchical lasso does not ``pay'' (in terms of
fitting) for those main effects, $\hbp_j-\hbn_j$, that are forced into
the model
by the hierarchy constraint to accommodate a strong interaction. Notice
that we
do pay for a nonzero main effect if \emph{both} $\hbp_j$ and $\hbn_j$
are nonzero. This makes sense since the constraint
could be satisfied with just one of these variables nonzero, but in
this case it
is advantageous to the fit to make both nonzero. In Figure \ref
{fig:df}, we
find that this bound is in expectation visually indistinguishable from
$E[\widehat{df}_{\lambda}]$.

\section{Related work}
\label{sec:related-work}
There has been considerable interest in fitting interaction models in
statistics and related fields. We focus here on an overview of
methods that aim at forming predictive models that satisfy the
hierarchical interactions restriction.

\subsection{Multi-step procedures}
\label{sec:multi-step-proc}
Many statistics textbooks discuss a simple stepwise procedure in which one
iteratively considers adding or removing the ``best'' variable
(whether it be main effect or interaction); they add that one should
only consider including an interaction if its main effects are in the
model [e.g., see backward elimination in \citet{Agresti02}, Section~6.1.3].
In doing so, they
are enforcing the strong hierarchy restriction. Such procedures are
ubiquitous [\citet{Nelder97},
\citet{Peixoto87}] as are more recent versions
[\citet{Friedman91},
\citet{Bickel08},
\citet{Park08},
\citet{Wu10}]. Another approach is to perform
model selection first without considering hierarchy and then to
include any lower-order terms necessary to satisfy hierarchy as a
post-processing step [\citet{Nardi07}]. Finally, \citet{Turlach04} and
\citet{Yuan07b}
consider modifying the LARS algorithm [\citet{LARS}] so that hierarchy
is enforced.

\subsection{Bayesian approaches}
\label{sec:bayesian-approach}
Another set of procedures for building hierarchical interaction
models comes from a Bayesian viewpoint. \citet{Chipman96} adapts the
\emph{stochastic search variable selection} (SSVS) approach of
\citet{George93} to produce strong or weak hierarchical interaction
models. SSVS makes use of a hierarchical normal mixture model to
perform variable selection in regression. Every variable has a latent
binary variable indicating whether it is ``active.'' Conditional on
this latent variable, each coefficient is a 0-mean normal with
variance determined by the latent importance of the coefficient. The
original SSVS paper chooses a prior in which the importance of each
variable is an independent Bernoulli. \citet{Chipman96} introduces
dependence into the prior so that $\Theta_{jk}$ is important
only if $\beta_j$ and/or $\beta_k$ is important as well.

\subsection{Optimization-based approaches}
\label{sec:optim-based-appr}
\citet{Choi10} formulate a nonconvex optimization problem to get sparse
hierarchical interaction models. They write\vadjust{\goodbreak}
$\Theta_{jk}=\Gamma_{jk}\beta_j\beta_k$, where $\beta$ are the main
effect coefficients and then apply $\ell_1$ penalties on $\beta$ and
$\Gamma$. Notice that $\Theta_{jk}\neq0$ implies $\beta_j\neq0$ and
$\beta_k\neq0$. The nonconvexity arises in writing $\Theta_{jk}$ as
the product of optimization variables.

Most similar to this paper's proposal is a series of methods which
formulate convex optimization problems to give sparse hierarchical
interaction models. \citet{Yuan09} modify the nonnegative
garrote [\citet{Breiman95}] by adding linear inequality constraints to
enforce hierarchy. In this sense, our method can be seen as the
adaption of
their approach to the lasso.

Finally, as discussed in Section~\ref{sec:sparsity-lasso}, another set
of convex methods makes use of the group lasso penalty
[\citet{Yuan06}]. \citet{Zhao09} [and, relatedly, \citet{Jenatton11}]
describe composite absolute
penalties (CAP), a very broad class of penalties that can achieve group and
hierarchical sparsity. To achieve ``hierarchical selection,'' they
put forward the principle that a penalty of the form
$\Vert(\phi_1,\phi_2)\Vert_\gamma+|\phi_1|$, with $\gamma>1$,
induces $\phi_2$ to be zero only when $\phi_1$ is zero as well. For
hierarchical interaction models, they suggest a penalty of the form
$\lambda\sum_{j<k} [|\Theta_{jk}|+\Vert(\beta_j,\beta
_k,\Theta
_{jk})\Vert_{\gamma_{j,k}} ]$.
This framework has been developed in the structured
sparsity literature [e.g., \citet{Bach12}]. \citet{Radchenko10}
introduce VANISH, which uses this nested-group principle to
achieve hierarchical sparsity in the context of nonlinear
interactions. Their penalty in the setting of \eqref{eq:model} is
$\sum_{j} [\lambda_1\Vert(\beta_j,\Theta_j)\Vert_2+\lambda
_2\Vert
\Theta_j\Vert_1 ]$.
As noted in Remark~\ref{re3}, our proposal is closer to CAP and VANISH than it
may first appear. Our problem can be rewritten to have a penalty of
the form $\lambda\sum_{j} [\max\{|\beta_j|,\Vert\Theta
_j\Vert_1\}
+(1/2)\Vert\Theta_j\Vert_1 ].$
In this sense, the penalty is in the spirit of CAP and related methods
although it
does not quite fall into the class
of CAP (since ours involves a sum of norms of norms). It is most
similar to VANISH in that it combines all of $\Theta_j$ into the
term involving $\beta_j$.

\section{Empirical study}
\label{sec:empirical-study}

\subsection{Simulations}
\label{sec:simulation}
Our main interest in this section is to study the advantages and
disadvantages of restricting one's interaction models to those that
honor hierarchy. Clearly, the effectiveness of such a strategy
depends on the true model generating the data. We take $n=100$ and
$p=30$ (435 two-way interactions) and consider
four scenarios:
\begin{longlist}[(III)]
\item[(I)] Truth is hierarchical: $\Theta_{jk}\neq0\implies\beta_j\neq
0, \beta_k\neq0$;
\item[(II)] Truth is anti-hierarchical: $\Theta_{jk}\neq0\implies\beta
_j=0, \beta_k=0$;
\item[(III)] Truth only has interactions: $\beta_{j}=0$ for all $j$;
\item[(IV)] Truth only has main effects: $\Theta_{jk}=0$ for all $jk$.
\end{longlist}
In cases (I), (II), (IV), we set 10 elements of $\beta$ to be nonzero (with
random sign), and, in cases (I), (II), (III), we set 20 elements of the
submatrix of $\Theta=\Theta^T$ to be nonzero.\vadjust{\goodbreak} The signal-to-noise
ratio (SNR) for the main effects part of the signal is about 1.5
whereas the SNR for the interactions part is about 1.

We study the effectiveness of the hierarchy constraint in the context
of both the lasso and forward stepwise
regression. Forward stepwise regression refers to a greedy strategy
for generating a sequence of linear regression models in which we start
with an intercept-only
model and then at each step add the variable that leads to the
greatest decrease in the residual sum of squares. We choose forward
stepwise as a basis of comparison since it has a simple modification
that we think may be the hierarchical interactions
approach most commonly used by statisticians. The modification is to
restrict the
set of interactions that could be added at a given step to only those
between main effect variables currently in the model. A backward
stepwise version of this approach is suggested in \citet{Peixoto87}.

We compare six methods, corresponding to each cell of the following table:\vspace*{9pt}
\begin{center}
\begin{tabular}{@{\extracolsep{\fill}}lccc@{}}
\hline
& \textbf{Hierarchical} & \textbf{All-pairs} & \textbf{Main effects
only}\\
\hline
{Lasso} &HL (our method)&APL&MEL\\
{Fwd stepwise}&HF&APF&MEF\\
\hline
\end{tabular}\vspace*{9pt}
\end{center}

\begin{figure}

\includegraphics{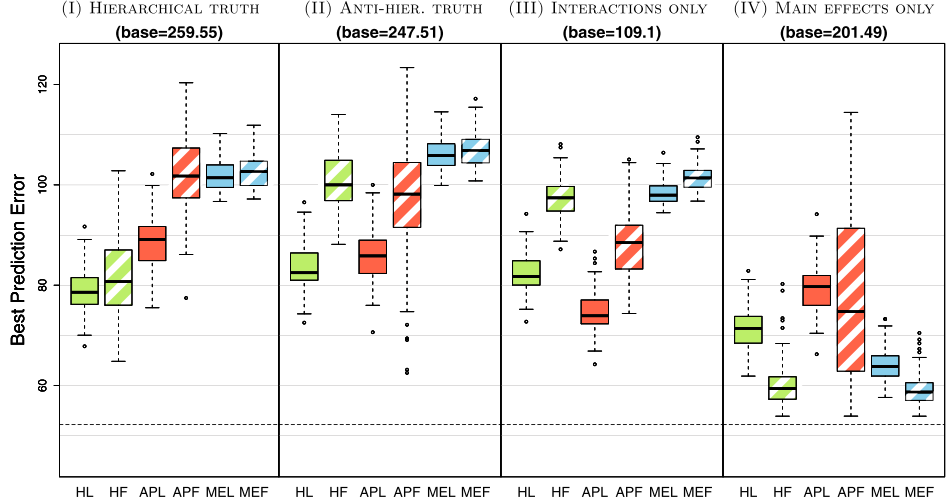}\vspace*{-3pt}

\caption{Prediction error: Dashed line shows Bayes error
(i.e., $\sigma^2$), and the base rate refers to the prediction error
of $\bar y_{\mathrm{train}}$. Green, red and blue colors indicate
hierarchy, all-pairs, and main effect only, respectively; solid
and striped indicate lasso and forward stepwise, respectively.}
\label{fig:sims-prederr}\vspace*{-3pt}
\end{figure}

Each method has a single tuning parameter: for
the lasso methods, the penalty parameter, $\lambda$, and for the
forward
stepwise methods, the number of variables, $k$. We fit each
method along a grid of tuning parameter values and select the model
with the smallest mean squared error, $E\Vert\hat
y-\mu\Vert^2$. Note that such an operation is only possible in
simulation since it requires knowing $\mu$; however, doing so avoids
the added variance of cross validation without being biased in favor
of any particular method. The results presented are
based on 100 simulations from the underlying model. Figure
\ref{fig:sims-prederr} shows the expected prediction error,
$\sigma^2+E[(\hat y-\mu)^2]$. Panel (I) shows that
%
%
%
%
\begin{figure}[b]\vspace*{-3pt}

\includegraphics{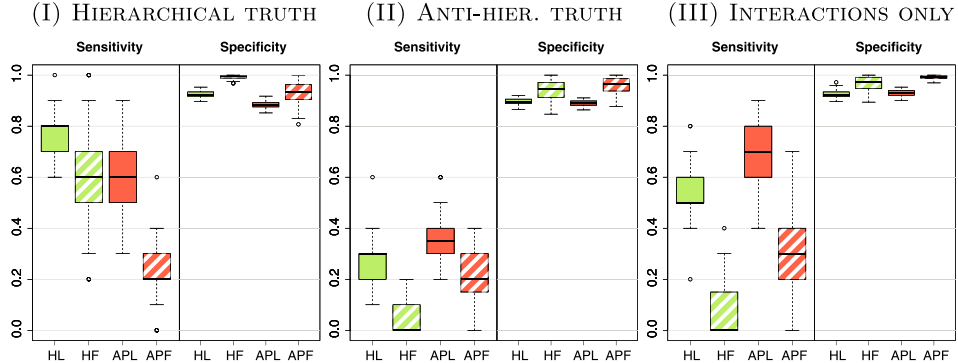}\vspace*{-3pt}

\caption{Plots show the ability of various methods to
correctly recover the nonzero interactions. This is the
sensitivity (i.e., proportion of $\Theta_{jk}\neq0$ for which
$\hTheta_{jk}\neq0$) and specificity (i.e., proportion of
$\Theta_{jk}=0$ for which $\hTheta_{jk}=0$) corresponding to the lowest
prediction error model of each method.}
\label{fig:sims-recov}
\end{figure}
when the truth is hierarchical, methods that assume hierarchy (HL, HF) do
better than the rest. These methods have ``concentrated'' their power
on the correct set of models and therefore receive the biggest payoff
for being correct. APL does better than MEL and MEF since it succeeds
in incorporating some of the correct interactions (recall that
interactions make up one quarter of the signal). In panel (II), we
notice our first surprise---that HL predicts well relative to the
others even when the truth is not hierarchical! We would have
expected APL (or APF) to be the clear winner in this situation since
surely the hierarchy assumption can only be detrimental in this
``anti-hierarchical'' scenario. The reason APL does not outperform HL
in this scenario is because APL has trouble identifying the main
effects (it gets swamped by the 435 interaction variables). In light
of Section~\ref{sec:effect-constraint}, this is where the hierarchy
constraint helps---main effects are penalized by less and interactions
by more. Even though APL is better able to find the correct
interactions than HL, as seen in panel (II) of Figure~\ref{fig:sims-recov},
APL does not predict as well as HL because it fails to find the main
effects, which constitute three quarters of the signal. Relatedly, in a
``hierarchical truth'' scenario
similar to (I) but with $p>n$ (not presented here), we have in fact observed
MEL doing better than APL (though not as well as HL) since APL is not
able to detect interactions accurately enough to make up for its inferior
ability to detect main effects. By contrast, HL does best in that
scenario, aided by hierarchy to capture both the main effect and
interaction components of the signal.\looseness=-1\vadjust{\goodbreak}

In panel (III), we see a situation where APL does dominate HL.
Since there are no main effects in the signal, all that is relevant is
a method's ability to find the interactions. HL identifies fewer
correct interactions than APL since any main effect ``information'' that
HL is using is spurious. Finally in panel (IV), we see a situation
where MEF, HF, MEL do better than the rest. Here again we find that
the hierarchy methods beat the
all-pairs methods since they favor main effects.

It is particularly illuminating to note the difference in
performance between HL and HF. HF in scenarios (II), (III) and (IV)
performs very similarly to the main effect only models. In (II) and
(III), HL does much better than HF both in terms of prediction error
and in ability to correctly identify interactions. HL appears to be
far less sensitive to violations of hierarchy than HF. This
difference is attributable to the joint nature in which HL acts: the
decision to include a main effect is made at the same time as
decisions about interactions. This allows a strong interaction to
``pull'' itself into the model. By contrast, HF selects main effects
with no regard to the information contained in the interactions.

%
%

\subsection{Data examples}
\label{sec:data-examples}
\begin{figure}

\includegraphics{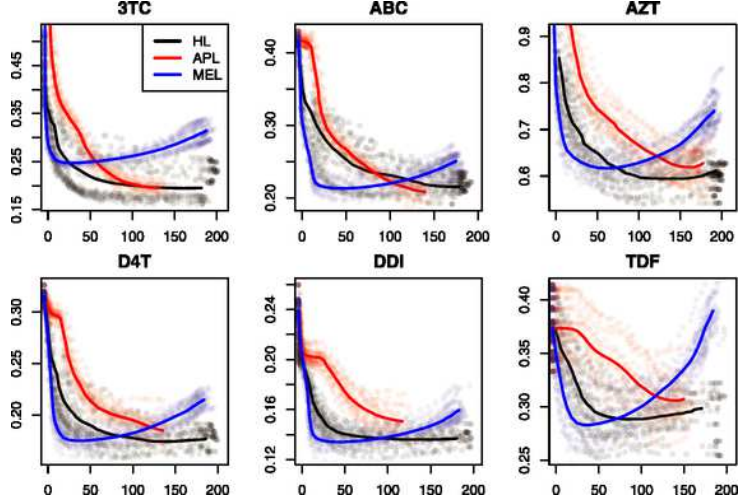}

\caption{HIV drug data: Test-set RMSE versus
practical sparsity (i.e., number of measured variables required
for prediction) for six different drugs. For each method,
the data from all 20 runs are displayed in faint colors; the thick lines
are averages over
these runs.}
\label{fig:nrti-rmse}
\end{figure}

Rhee et~al. (\citeyear{Rhee06}) study six nucleoside reverse transcriptase inhibitors
(NRTIs) that are used to treat HIV-1. The target of these drugs can
become resistant through mutation, and \citet{Rhee06} compare a
collection of models for predicting these drug's (log)
susceptibility---a~measure of
drug resistance---based on the location of mutations. In the six
cases, there are between $p=211$ and $p=218$ sites
with mutations occurring in the $n=784$ to $n=1073$ samples. While
they focus on main effect only models, we consider here the all-pairs
lasso (APL)
and weak hierarchical lasso (HL) in addition to the standard main
effects lasso
(MEL). We train on half of the samples and test on the remaining
samples. To reduce the dependence of the results on the particular
random training-test split, we repeat this process twenty times and
average the results.
Figure~\ref{fig:nrti-rmse} shows the average test RMSE versus the
average practical sparsity for each of the six
drugs. In all cases but ABC, we find that HL achieves a better test
error at most levels of practical sparsity than APL. That said, if the
number of
mutations one has to measure is not of concern (so that we can choose
for each method the minimum RMSE model), then no method dominates in
all the situations. It is worth conceding---since this is a paper on
interactions---that in several of the cases a pure main effects model
appears to be the best option.

\section{Algorithmics}
\label{sec:algorithmics}
Some of the fastest lasso solvers rely on coordinate descent,
which amounts to iteratively applying \eqref{eq:lasso} until
convergence [\citet{glmnet}]. \citet{Tseng01} proves that blockwise
coordinate descent converges
to the global minimum for a convex problem specifically when the\vadjust{\goodbreak}
nondifferentiable part of
the problem is blockwise separable. In the case of the strong
hierarchical lasso, the hierarchy constraints
combined with the symmetry constraint couple all the parameters
together, meaning that coordinate descent is prone to getting stuck at
suboptimal points. To see this, note that $\Theta_{jk}=\Theta_{kj}$ appears
in two constraints $\Vert\Theta_j\Vert_1\le\bp_j+\bn_j$ and
$\Vert\Theta_k\Vert_1\le\bp_k+\bn_k$. By contrast, the
constraints in
the weak hierarchical lasso problem are blockwise separable so that
blockwise coordinate
descent on blocks of the form $(\Theta_j,\bp_j,\bn_j)$ for
$j=1,\ldots,p$ does work. We begin by discussing our approach to
solving the weak hierarchical lasso problem. In Section
\ref{sec:admm} we discuss how we can solve a sequence of weak
hierarchical lasso problems that converges to a solution of the strong
hierarchical lasso.

\begin{algorithm}[t]
\caption{\texttt{WEAK-HIERNET}: Generalized gradient descent to solve
weak hierarchical lasso,
\protect\eqref{eq:weak-hiernet}, with elastic net penalty $\varepsilon$.}\label{alg:gg}
\emph{Inputs:} $X\in\real^{n\times p}, Z\in\real^{n\times
p(p-1)}, \lambda>0$. Initialize $(\hat\beta^{+(0)},\hat\beta
^{-(0)},\widehat\Theta^{(0)})$.\\
For $k=1,2, \ldots$ until convergence:

Compute residual: $\hat r^{(k-1)}\leftarrow y-X(\hat\beta
^{+(k-1)}-\hat\beta^{-(k-1)})-Z\hTheta^{(k-1)}/2.$

For $j=1,\ldots,p$:
\begin{eqnarray*}
\bigl(\hat\beta^{+(k)}_j,\hat\beta^{-(k)}_j,
\widehat\Theta ^{(k)}_j\bigr)&\leftarrow&\texttt{ONEROW}
\bigl(\delta\hat\beta ^{+(k-1)}_j-tX_j^T
\hat r^{(k-1)},
\\
&&\hspace*{45pt}\delta\hat\beta^{-(k-1)}+tX_j^T\hat
r^{(k-1)},
\\
&&\hspace*{45pt}\delta\widehat\Theta^{(k-1)}_j-tZ_{(j,\cdot)}^T
\hat r^{(k-1)}\bigr),
\end{eqnarray*}
where \texttt{ONEROW} is given in Algorithm~\ref{alg:onerow}, $\delta=1-t\varepsilon$, and
$Z_{(j,\cdot)}\in\real^{n\times(p-1)}$ denotes the columns of $Z$
involving $X_j$.
\end{algorithm}

\subsection{Solving the weak hierarchical lasso}
\label{sec:solv-weak-hier}

While blockwise coordinate descent would work for solving the weak
hierarchical lasso problem,
we instead describe a generalized gradient descent approach. Given a
problem of the form
%
\begin{equation}
\label{eq:gg-problem} \mathop{\Minimize}_\phi g(\phi) + h(\phi),
\end{equation}
in
which $g$ is convex and differentiable with a Lipschitz gradient and
$h$ is convex, generalized
gradient descent works by solving a sequence of problems of the form
\[
\hat\phi_k\leftarrow\mathop{\arg\min}_\phi
\frac{1}{2t}\bigl\Vert\phi- \bigl[\hat\phi_{k-1}-t\nabla g(\hat
\phi_{k-1})\bigr]\bigr\Vert^2+h(\phi),
\]
where $t$ is a suitably chosen step size [\citet{Beck09}]. These
subproblems are easier to solve than \eqref{eq:gg-problem} since they
replace $g$ by a spherical quadratic. Under the previously stated
conditions, generalized gradient descent is
guaranteed to get within $O(1/k)$ of the optimal value after $k$
steps; in fact, with a simple modification to the
algorithm, this rate improves to $O(1/k^2)$ [\citet{Beck09}]. Looking
back at
\eqref{eq:weak-hiernet}, we take $g$ to be the differentiable part,
$q(\beta_0,\bp-\bn,\Theta)+\lambda1^T(\bp+\bn)$ and $h$ to be the
$\ell_1$ penalty on $\Theta$ and the set of constraints. The
subproblem is of the form
\begin{eqnarray*}
&&\mathop{\Minimize}_{\beta^\pm\in\real^p, \Theta\in\real^{p\times
p}} \frac
{1}{2t}\bigl\Vert\bp-\tilde\beta^+
\bigr\Vert^2+\frac{1}{2t}\bigl\Vert\bn -\tilde\beta ^-\bigr\Vert^2+
\frac{1}{2t}\Vert\Theta-\widetilde\Theta\Vert^2_F +
\frac
{\lambda}{2}\Vert\Theta\Vert_1
\\
&&\qquad\mathrm{s.t.} \qquad %
\left.\begin{aligned} \Vert\Theta_j
\Vert_1&\le\bp_j+\bn_j
\\
\bp_j&\ge0, \bn_j\ge0
\\
\end{aligned} %
\right\}\mbox{ for }j=1,\ldots,p,
\end{eqnarray*}
where $(\tilde\beta^+,\tilde\beta^-,\widetilde\Theta)$ depends on the
previous iteration's solution and the data, $X$, $Z$ and $y$. The
exact form of $(\tilde\beta^+,\tilde\beta^-,\widetilde\Theta)$ is
given in
Algorithm~\ref{alg:gg} for solving~\eqref{eq:weak-hiernet}, where
$q(\beta_0,\beta,\Theta)$ includes an elastic net penalty as
described in Remark~\ref{rem2} of Section~\ref{sec:our-proposed-method}.
The above problem
decouples into $p$ separate pieces involving $(\Theta_j,\bp_j,\bn_j)$
that could be solved in
parallel:
%
\begin{eqnarray}
\label{eq:prox}\qquad&& \mathop{\Minimize}_{\beta_j^\pm\in\real, \Theta_j\in\real^{p-1}} \frac
{1}{2t}\bigl(
\bp_j-\tilde\beta^+_j\bigr)^2+
\frac{1}{2t}\bigl(\bn_j-\tilde\beta ^-_j
\bigr)^2+\frac{1}{2t}\Vert\Theta_j-\widetilde
\Theta_j\Vert^2 +\frac
{\lambda}{2}\Vert
\Theta_j\Vert_1
\nonumber
\\[-8pt]
\\[-8pt]
\nonumber
&&\qquad\mbox{s.t.}\qquad \Vert\Theta_j\Vert_1\le
\bp_j+\bn_j, \bp_j\ge 0, \bn_j
\ge 0.
\end{eqnarray}
In Appendix~\ref{app:prox}, we derive
an algorithm, \texttt{ONEROW}, that solves \eqref{eq:prox} based on the
observation
that, in terms of an optimal dual variable $\hat\alpha$, a solution is simply
$\hTheta_j=\mathcal S[\widetilde\Theta_j, t(\lambda/2+\hat\alpha)]$
and $\hat\beta^\pm_j=[\tilde\beta^\pm_j+t\hat\alpha]_+$.

We solve \eqref{eq:weak-hiernet} along a sequence of $\lambda$
values, from large to small, using the solution from the previous
$\lambda$ as a warm
start for the next. The \texttt{WEAK-HIERNET} algorithm gets within
$\varepsilon$ of the optimal value of \eqref{eq:weak-hiernet} in
$O(p^2\max\{n,p\}/\varepsilon)$ time.

\subsection{Solving the strong hierarchical lasso}
\label{sec:admm}
In Section~\ref{sec:solv-weak-hier}, we noted that each step of
generalized gradient descent conveniently decouples into $p$
single-variable optimization problems. However, for the strong
hierarchical lasso, \eqref{eq:hiernet}, the symmetry
constraint ties all variables together. We therefore make use of
Alternating Direction Method of Multipliers (ADMM), which is a very
widely applicable framework that allows convex problems to be split
apart into separate easier subproblems [\citet{Boyd10}].

Given a convex problem of the form $\Minimize_{\phi}f(\phi)+g(\phi)$,
we rewrite
it equivalently as $\Minimize_{\phi,\varphi}f(\phi)+g(\varphi
)\mbox{
s.t. }\phi=\varphi$, and then the ADMM algorithm repeats the
following three steps until convergence:
\begin{longlist}[(1)]
\item[(1)]
$\hat\phi=\argmin_{\phi} [f(\phi)+(\rho/2)\Vert\phi-\hat
\varphi
+\hat u/\rho\Vert^2 ]$.
\item[(2)]
$\hat\varphi=\argmin_{\varphi} [g(\varphi)+(\rho/2)\Vert
\varphi
-\hat\phi-\hat u/\rho\Vert^2 ]$.
\item[(3)] $\hat u \leftarrow\hat u + \rho(\hat\phi-\hat\varphi)$.
\end{longlist}
Thus the ADMM algorithm separates the two difficult parts of the
problem, $f$
and~$g$, into separate optimization problems. The dual variable $\hat
u$ serves to pull these two problems together, resulting in an
algorithm that is guaranteed to converge to a solution as long as
$\rho>0$. In practice, the value of $\rho$ affects the speed of convergence.

In our case, we use ADMM to separate the hierarchy constraints,
involving $(\bp,\bn,\Theta)$ from
the symmetry constraint, which will involve a symmetric version of
$\Theta$, which we
call $\Omega$:
%
\begin{eqnarray}
\label{eq:admm-problem} &&\mathop{\operatorname{Minimize}\limits_{\beta_0\in\real, \beta^\pm\in\real
^p,}}\limits_{\Theta,\Omega\in
\real^{p\times p}} q\bigl(
\beta_0,\bp-\bn,\Theta\bigr) + \lambda1^T\bigl(\bp+\bn\bigr) +
\frac{\lambda}{2}\Vert\Theta\Vert_1
\nonumber
\\[-8pt]
\\[-8pt]
\nonumber
& &\qquad\mbox{s.t.} \qquad\Omega= \Omega^T,\Theta=\Omega %
\left.\begin{aligned} \Vert\Theta_j\Vert_1&\le
\bp_j+\bn_j
\\
\bp_j&\ge0, \bn_j\ge0
\\
\end{aligned} %
\right\} \mbox{ for }j=1,\ldots,p.
\end{eqnarray}
The resulting ADMM algorithm is given in Algorithm~\ref{alg:admm},
which is explained in greater detail in Section~4 of the
supplementary materials [\citet{Bien13supp}]. Conceptually,
the algorithm alternately updates two matrices, $\Theta$ and~$\Omega$.
Throughout the algorithm, we update $\Theta$ by solving a version of
problem~\eqref{eq:weak-hiernet}, and we update $\Omega$ by
symmetrizing a version of $\Theta$. At convergence,
$\hTheta=\widehat\Omega$, and thus $\hTheta$ is both symmetric and
satisfies the hierarchy constraints.
\begin{algorithm}[t]
\caption{\texttt{STRONG-HIERNET}: Solve \protect\eqref{eq:hiernet}
via ADMM.}\label{alg:admm}
\emph{Inputs:} $X\in\real^{n\times p}, Z\in\real^{n\times
p(p-1)}, \lambda>0, \rho>0$.\vspace*{2pt}

Initialize
$(\hat\beta^+,\hat\beta^-,\hTheta), \widehat\Omega, \widehat
U$.\vspace*{2pt}

Repeat until convergence:
\begin{enumerate}
\item[(1)] \texttt{WEAK-HIERNET}($X,Z,\lambda$), but in the call to
\texttt{ONEROW} replace
the argument $\delta\hTheta^{(k-1)}_j-tZ_{(j,\cdot)}^T\hat
r^{(k-1)}$ with
$\delta\hTheta^{(k-1)}_j-tZ_{(j,\cdot)}^T\hat
r^{(k-1)}+\rho[\hTheta^{(k-1)}_j-\widehat\Omega_j]+\widehat U_j$. Also,
initialize with $(\hat\beta^+,\hat\beta^-,\hTheta)$.
\item[(2)]

$\widehat\Omega\leftarrow\frac{1}{2}(\hTheta+\hTheta^T)+\frac
{1}{2\rho
}(\widehat U+\widehat U^T)$.
\item[(3)] $\widehat U\leftarrow\widehat U + \rho(\hTheta-\widehat
\Omega)$.
\end{enumerate}
\end{algorithm}

\section{Discussion}
\label{sec:discuss}
In this paper, we have proposed a modification to the lasso for fitting
strong and weak hierarchical interaction models. These two approaches
are closely tied, and our algorithms to solve the two exploit their
similar structure. A~key advantage of our framework is that it admits
a simple characterization of the effect of imposing hierarchy.
We compare our hierarchical methods to the lasso and to stepwise
procedures to understand the implications of demanding hierarchy.
We introduce a distinction between models that
have a small number of parameters and those that require measuring
only a small number of variables. The hierarchical interaction
requirement favors models with the latter type of sparsity, a feature
that is desirable when performing measurements is costly, time
consuming, or otherwise inconvenient. The \texttt{R} package
\texttt{hierNet} provides implementations of our strong and weak
methods, both for
Gaussian and logistic losses. This work has potential applications to
genomewide association studies. In future work, we intend to extend
this framework to contexts in which only certain interactions should
be considered such as in gene-environment interaction models.

\begin{appendix}\label{app}

\section{Proofs of strong and weak hierarchy}
\label{app:proof-hier}

We begin by proving Lemma 1, which characterizes all solutions to
\eqref{eq:hiernet} as a relatively simple function of $y$. The
structure of our proof is based on (Ryan) \citeauthor{RyanTibs11} (\citeyear{RyanTibs11,RyanTibs12}).
\subsection{Characterizing the solution}
\label{sec:char-solut}

For ease of analysis, we write \eqref{eq:hiernet} equivalently in
terms of $\Theta^+$ and $\Theta^-$. Also, for notational simplicity, we
write $\phi=(\bp,\bn,\Theta^+,\Theta^-)$ and
$\tX=(X;-X;Z/2;-Z/2)$. The strong hierarchical lasso
problem is the following:
\begin{eqnarray*}
&&\mathop{\Minimize}_{\bp,\bn,\Theta^+,\Theta^-} \frac{1}{2}\Vert y-\tX\phi \Vert^2 +
\lambda_11^T\bigl(\bp+\bn\bigr)+\lambda_2\bigl
\langle11^T,\Theta^++\Theta ^-\bigr\rangle
\\
&&\qquad\mbox{s.t.} \qquad 1^T\bigl(\Theta^+_{j}+\Theta^-_{j}
\bigr) \le \bp_j+\bn_j\mbox{ and }\bp_j
\ge0, \bn_j\ge0\mbox{ for each $j$},
\\
&&\hspace*{59pt}\Theta^+-\Theta^-=\Theta^{+T}-\Theta^{-T},
\Theta_{jk}^\pm \ge 0, \Theta_{jj}^\pm=0.
\end{eqnarray*}
In introducing $\Theta^\pm$, we are
not in fact changing the problem since at a solution
$\hTheta^\pm=\max\{\pm\hTheta,0\}$ (for $\lambda_2>0$). To see this,
note that given any feasible point with $\Theta_{jk}^+>0$ and $\Theta
_{jk}^->0$,
we can produce a feasible point with strictly lower objective by
reducing $\Theta_{jk}^+,\Theta_{jk}^-,\Theta_{kj}^+,\Theta_{kj}^-$ all
by equal amounts.

We will try to make this as close as possible in form and notation to (Ryan)
\citeauthor{RyanTibs12}'s (\citeyear{RyanTibs12}) treatment of the
generalized lasso problem. Our
optimization problem is of the form
%
\begin{equation}
\label{eq:linear} \mathop{\Minimize}_{\phi} \frac{1}{2}\Vert y-\tX\phi
\Vert^2 + w^T\phi\qquad \mbox{s.t.}\qquad  D\phi\ge0, L\phi=0.
\end{equation}
The Lagrangian of this problem is
\[
L(\phi;\mu,\nu)=\frac{1}{2}\Vert y-\tX\phi\Vert^2 +
w^T\phi-\mu^TD\phi+\nu^TL\phi,
\]
where $\mu\ge0$ and $\nu$ are dual variables. The KKT conditions for
$(\hat\phi(y),(\hat\mu(y), \hat\nu(y)))$ to be an optimal primal-dual
pair are the following:
\begin{eqnarray*}
\tX^T(y-\tX\hphi)&=&w-D^T\hmu+L^T\hnu,
\\
\hmu_i(D\hphi)_i&=&0,
\\
\hmu&\ge&0,
\\
D\hphi&\ge&0,\qquad L\hphi=0.
\end{eqnarray*}
Now, define the ``boundary'' and ``active'' sets as
\begin{eqnarray*}
\B(\hmu)&=&\{i\dvtx \hmu_i=0\},
\\
\A(\hphi)&=&\bigl\{i\dvtx [D\hphi]_i>0\bigr\}.
\end{eqnarray*}
These are not necessarily unique since
$(\hphi,(\hmu,\hnu))$ may not be unique. In terms of the active set
$\A(\hphi)$, the KKT conditions become
\begin{eqnarray*}
\tX^T(y-\tX\hphi)&=&w-D^T_{-\A(\hphi)}
\hmu_{-\A(\hphi)}+L^T\hnu,
\\
\hmu_{\A(\hphi)}&=&0, \qquad \hmu\ge0,
\\
L\hphi&=&0,\qquad D_{-\A(\hphi)}\hphi=0.
\end{eqnarray*}
Solving
for $\hphi$, we get the following characterization of a strong
hierarchical lasso solution:

\begin{lemma}\label{lem:project}
Suppose $\hphi$ is a solution to the strong hierarchical lasso problem
\eqref{eq:hiernet} [taking $q(\beta_0,\beta,\Theta)$ to be the\vadjust{\goodbreak}
quadratic loss] with $\A(\hphi)=\{i\dvtx [D\hphi]_i>0\}.$ Then,
$\hphi$ can be written in terms of $\A(\hphi)$ and $y$ as
\[
\hphi=(\tX P_{\ns(L)\cap\ns(D_{-\A(\hphi)})})^+\bigl(y-\bigl(P_{\ns(L)\cap
\ns
(D_{-\A(\hphi)})}\tX^T
\bigr)^+w\bigr)+b,
\]
where $b\in\ns(\tX)\cap\ns(L)\cap\ns(D_{-\A(\hphi)})$ satisfies
\[
D_{i}\bigl[(\tX P_{\ns(L)\cap\ns(D_{-\A(\hphi)})})^+\bigl(y-\bigl(P_{\ns(L)\cap\ns(D_{-\A
(\hphi
)})}
\tX^T\bigr)^+w\bigr)+b\bigr]>0
\]
for all $i\in\A(\hphi)$.\vspace*{-3pt}
\end{lemma}
\begin{pf}
Defining $\tD=
{
D_{-\A(\hphi)}\choose L}
$ and $P=P_{\ns(\tD)}=P_{\ns(L)\cap\ns(D_{-\A(\hphi)})}$, we
solve for
$\hphi$ in the same manner as is done in (Ryan)
\citet{RyanTibs12}. Since $\tD\hphi=0$ is equivalent to $P\hphi
=\hphi$,
we have $P\tX^T(y-\tX P\hphi)=Pw$. We see that
$Pw\in\operatorname{col}(P\tX^T)$ and thus $Pw=(P\tX^T)(P\tX^T)^+Pw$. Thus,
$P\tX^T\tX P\hphi=P\tX^T(y-(P\tX^T)^+Pw)$ from which we get
\[
\hphi=(\tX P)^+\bigl(y-\bigl(P\tX^T\bigr)^+Pw\bigr)+b
\]
for $b\in\ns(\tX P)$ and such that $\tD b=0$ and $D_i\hphi>0$ for
$i\in\A(\hphi)$. To complete the result, we observe that the first two
conditions reduce to
$b\in\ns(\tX)\cap\ns(L)\cap\ns(D_{-\A(\hphi)})$.\vspace*{-3pt}
\end{pf}

We will use this characterization of a solution both to prove that
the hierarchy property holds with probability one under weak
assumptions and to derive an unbiased estimate of the degrees of freedom.

Before we do so, we write out $\tD={
D_{-\A(\hphi)}\choose L
}$ more explicitly and introduce
a little notation that will be useful later.
Every row of $D$ corresponds to an inequality constraint, and we can
describe these rows in terms of ten subsets,
%
\begin{eqnarray}
\label{eq:sets} \L&=&\bigl\{j\dvtx 1^T\bigl(\hTheta^+_{j}+
\hTheta^-_{j}\bigr) < \hbp_j+\hbn_j\bigr\},\qquad
\T=\L ^c,
\nonumber
\\[-2pt]
\P\bigl(\hat\beta^\pm\bigr)&=&\bigl\{j\dvtx \hat\beta^\pm_j>0
\bigr\},\qquad \Z\bigl(\hat\beta ^\pm\bigr)=\P \bigl(\hat
\beta^\pm\bigr)^c,
\\[-2pt]
\P\bigl(\hTheta^\pm\bigr)&=&\bigl\{j\neq k\dvtx \hTheta^\pm_{jk}>0
\bigr\},\qquad \Z\bigl(\hTheta ^\pm \bigr)=\P\bigl(\hTheta^\pm
\bigr)^c.
\nonumber
\end{eqnarray}
The set $\A(\hphi)^c$ is made up of $\T$, $\Z(\hbp)$, $\Z(\hbn)$,
$\Z(\hTheta^+)$ and $\Z(\hTheta^-)$. The matrix $\tD$ has
$2p+2p^2$ columns
that can be partitioned as
$(\tD^{\bp}\dvtx \tD^{\bn}\dvtx \tD^{\Theta^+}\dvtx \tD^{\Theta^-})$
and a row for every constraint. The rows of this matrix are the
following (where $e_j$ and~$1_p$ are row vectors):

\tabcolsep=0pt
\begin{tabular}{l@{\hspace*{10pt}}lc@{\hspace*{8pt}}c@{\hspace*{8pt}}c@{\hspace*{8pt}}ccl}
R1. For each $j\in\T$, & ( & $e_j$ & $e_j$ & $-e_j\otimes1_p$ &
$-e_j\otimes1_p$&)\\
R2. For each $j\in\Z(\bp)$, & ( & $e_j$ & 0 & 0 & 0&)\\
R3. For each $j\in\Z(\bn)$, & ( & 0 & $e_j$ & 0 & 0&)\\
R4. For each $jk\in\Z(\Theta^+)$, & ( & 0 & 0 & $e_j\otimes e_k$ &
0&)\\
R5. For each $jk\in\Z(\Theta^-)$, & ( & 0 & 0 & 0 & $e_j\otimes
e_k$&)\\
R6. For each $j$, & ( & 0 & 0 & $e_j\otimes e_j$ & 0&)\\
R7. For each $j$, & ( & 0 & 0 & 0 & $e_j\otimes e_j$&)\\
R8. For each $j<k$, & ( & 0 & 0 & $e_j\otimes e_k+e_k\otimes e_j$ &
$-e_j\otimes e_k-e_k\otimes e_j$ & ).   \vadjust{\goodbreak}
\end{tabular}

We will refer to this in the proofs that follow.

\subsection{\texorpdfstring{Proof of strong hierarchy (Property \protect\ref{prop:strong-hier})}
{Proof of strong hierarchy (Property 2)}}
\label{sec:prov-strong-hier}
Including the elastic net penalty,
$(\varepsilon/2)\Vert\bp\Vert^2+(\varepsilon/2)\Vert\bn\Vert
^2+(\varepsilon
/2)\Vert\Theta\Vert^2_F$,
is equivalent to replacing $\tX$ and $y$ in \eqref{eq:linear} by
\begin{eqnarray*}
\tX_\varepsilon=\pmatrix{ X&-X&Z/2&-Z/2\vspace*{2pt}
\cr
\sqrt{
\varepsilon}I_{p}&0&0&0\vspace*{2pt}
\cr
0&\sqrt{\varepsilon}I_{p}&0&0
\vspace*{2pt}
\cr
0&0&\sqrt{\varepsilon}I_{p^2}&-\sqrt{\varepsilon}I_{p^2}
} \quad\mbox{and}\quad y_\varepsilon= \pmatrix{ y\vspace*{2pt}
\cr
0_{(2p+p^2)\times1}}.
\end{eqnarray*}
Suppose we solve \eqref{eq:linear} with the above design matrix. By
Lemma~\ref{lem:project},
\[
\hphi=(\tX_\varepsilon P_{\ns(L)\cap\ns(D_{-\A(\hphi
)})})^+\bigl(y_\varepsilon-\bigl(P_{\ns
(L)\cap\ns(D_{-\A(\hphi)})}
\tX_\varepsilon^T\bigr)^+w\bigr)+b
\]
for some $b\in\ns(\tX_\varepsilon)\cap\ns(L)\cap\ns(D_{-\A(\hphi
)})$ satisfying
\[
D_{i}\bigl[(\tX P_{\ns(L)\cap\ns(D_{-\A(\hphi)})})^+\bigl(y_\varepsilon-\bigl(P_{\ns(L)\cap\ns(D_{-\A
(\hphi
)})}
\tX^T\bigr)^+w\bigr)+b\bigr]>0
\]
for all $i\in\A(\hphi)$. Let $S_{v}\dvtx \real^{2p+2p^2}\to\real
^{|v|}$ be the
linear operator that selects the part of a vector corresponding to
the variable $v$. Now, $b\in\ns(\tX_\varepsilon)$ implies
that $S_{\bp}(b)=S_{\bn}(b)=0$ and $S_{\Theta^+}(b)=-S_{\Theta^-}(b)$.
We showed earlier that we cannot have $\hTheta_{jk}^+>0$ and
$\hTheta_{jk}^->0$. This means that for any $jk$, there must be an
$i\notin\A(\hphi)$
for which $D_i\hphi=0$ corresponds to $S_{\Theta^+_{jk}}(\hphi)=0$ or
$S_{\Theta^-_{jk}}(\hphi)=0$. Thus, $D_{-\A(\hphi)}b=0$ means that
$S_{\Theta^+_{jk}}(b)=0$ or $S_{\Theta^-_{jk}}(b)=0$ for each $jk$.
This implies that
$\ns(\tX_\varepsilon)\cap\ns(D_{-\A(\hphi)})=\{0\}$ and thus $b=0$.

We show now that $P (\bigcup_j\{\hbp_j=\hbn_j>0\} )=0.$ In
terms of our above notation, this is
%
\begin{equation}
\label{eq:zeroprob} P \biggl(\bigcup_{j}\bigl\{
\hphi^T\bigl(e_j^+-e_j^-\bigr)=0,
\hphi^Te_j^+>0, \hphi ^Te_j^->0
\bigr\} \biggr)=0,
\end{equation}
where $e_j^\pm\in\real^{2p+2p^2}$ is the vector with all zeros except
for $S_{\beta^\pm}(e_j^\pm)=1$.
Let $P=P_{\ns(L)\cap\ns(D_{-\A})}$, and consider the set
\begin{eqnarray*}
\N&=&\bigcup_{\A}\bigcup_{j=1}^p\bigl\{z\dvtx \bigl[(
\tX_\varepsilon P)^+\bigl(z-\bigl(P\tX_\varepsilon^T\bigr)^+w
\bigr)\bigr]^T\bigl(e_j^+-e_j^-\bigr)=0,
\\
&&\hspace*{32pt}\bigl[(\tX_\varepsilon P)^+\bigl(z-\bigl(P\tX_\varepsilon^T
\bigr)^+w\bigr)\bigr]^Te_j^+>0,
\\
&&\hspace*{77pt}\bigl[(\tX_\varepsilon P)^+\bigl(z-\bigl(P\tX_\varepsilon^T
\bigr)^+w\bigr)\bigr]^Te_j^->0\bigr\}.
\end{eqnarray*}
In light of \eqref{eq:sets}, fixing $\A$ automatically specifies
$\T,\P(\hat\beta^\pm),\P(\hTheta^\pm)$. The outer union is restricted
to those subsets $\A$ of $2p+2p^2$ elements that
would have $\P(\hTheta^+)\cap\P(\hTheta^-)=\varnothing$ and
$\P(\hbp)\cap\P(\hbn)\subseteq\T$. The event\vadjust{\goodbreak}
in \eqref{eq:zeroprob} is contained in $\{y_\varepsilon\in\N\}$ since it
corresponds to
the case in which $\A$ is $\A(\hphi)$. We begin by showing that $\A
(\hphi)$ is in
this restricted union with probability one. We have already argued
that at a solution we must have $\hTheta_{jk}^+\hTheta_{jk}^-=0$ for
all $jk$. Now, $\hbp_j>0$ and $\hbn_j>0$ together imply that
$1^T(\hTheta^+_j+\hTheta^-_j)=\hbp_j+\hbn_j$ since otherwise we could
lower the objective by reducing $\hbp_j$ and $\hbn_j$ without leaving
the feasible set. Therefore, it would be sufficient
to show that $P(y_\varepsilon\in\N)=0$. We do so by observing that $\{
y_\varepsilon\in\N\}$
is a finite union of zero probability sets.\vspace*{1pt}

We begin by establishing that $\ns(I_{1:n}(P\tX_\varepsilon^T)^+)=\ns
(\tX
P)$.
To do so, we
write $P=UU^T$ for some
$U^TU=I$. Now, $\operatorname{row}(\tX P)\subseteq\operatorname{row}(P)=\operatorname{col}(U)$,
so we can
write $U=(U_1:U_2)$, where $\operatorname{row}(\tX P)=\operatorname{col}(U_1)$ and
thus $\tX PU_2=0$. Since $PU_2=U_2$, it follows that $\tX U_2=0$.
Write $U_i^T=(U_i^{\bp T}\dvtx U_i^{\bn
T}\dvtx U_i^{\Theta^+T}\dvtx\break U_i^{\Theta^-T})$ for $i=1,2$, and observe that
$U_i^T\tX_\varepsilon^T=[U_i^T\tX^T\dvtx \sqrt{\varepsilon}U_i^{\bp
T}\dvtx \sqrt{\varepsilon}U_i^{\bn T}\dvtx \break \sqrt{\varepsilon}(U_i^{\Theta
^+T}-U_i^{\Theta^-T})]$
so that
\begin{eqnarray*}
U_1^T\tX_\varepsilon^T
\tX_\varepsilon U_2&=&0+\varepsilon\bigl[U_1^{\bp
T}U_2^{\bp}+U_1^{\bn
T}U_2^{\bn}+
\bigl(U_1^{\Theta^+}-U_1^{\Theta^-}\bigr)
\bigl(U_2^{\Theta
^+}-U_2^{\Theta
^-}\bigr)\bigr]
\\
&=&\varepsilon\bigl[U_1^TU_2-U_1^{\Theta^+T}U_2^{\Theta^-}-U_1^{\Theta
^-T}U_2^{\Theta^+}
\bigr]
\\
&=&-\varepsilon\biggl[0+\sum_{jk}\bigl[U_1^{\Theta^+}
\bigr]_{jk}\bigl[U_2^{\Theta
^-}\bigr]_{jk}+
\bigl[U_1^{\Theta^-}\bigr]_{jk}\bigl[U_2^{\Theta^+}
\bigr]_{jk}\biggr].
\end{eqnarray*}
Now, for each $jk$ we must have $jk\in\Z(\Theta^+)\cup\Z(\Theta^-)$
since $\P(\Theta^+)\cap\P(\Theta^-)=\varnothing$. Thus, for each
$jk$, there
is a R4 or R5 row in $D_{-\A}$ and thus $D_{-\A}(U_1:U_2)=0$ implies
that $[U_1^{\Theta^+}]_{jk}=0$ or $[U_2^{\Theta^-}]_{jk}=0$ for each
$jk$ and likewise $[U_2^{\Theta^+}]_{jk}=0$ or
$[U_1^{\Theta^-}]_{jk}=0$. Therefore,
$U_1^T\tX_\varepsilon^T\tX_\varepsilon U_2=0$. Now,
\begin{eqnarray*}
I_{1:n}\bigl(P\tX_\varepsilon^T\bigr)^+&=&\tX P\bigl(P
\tX_\varepsilon^T\tX_\varepsilon P\bigr)^+ =\tX
UU^T\bigl(UU^T\tX_\varepsilon^T
\tX_\varepsilon UU^T\bigr)^+
\\
&=&\tX UU^TU\bigl(U^T\tX_\varepsilon^T
\tX_\varepsilon U\bigr)^+U^T =\tX U\bigl(U^T
\tX_\A^T\tX_\varepsilon U\bigr)^+U^T
\\
&=&\tX(U_1 U_2) \pmatrix{ U_1^T
\tX_\varepsilon^T\tX_\varepsilon U_1&U_1^T
\tX_\varepsilon^T\tX_\varepsilon U_2\vspace*{2pt}
\cr
U_2^T\tX_\varepsilon^T
\tX_\varepsilon U_1&U_2^T
\tX_\varepsilon^T\tX _\varepsilon U_2}
^+U^T
\\
&=& (\tX U_1 0) \pmatrix{ U_1^T
\tX_\varepsilon^T\tX_\varepsilon U_1&0
\vspace*{2pt}
\cr
0&U_2^T\tX_\varepsilon^T
\tX_\varepsilon U_2 }^+ U^T
\\
&=& \tX U_1\bigl(U_1^T\tX_\varepsilon^T
\tX_\varepsilon U_1\bigr)^+U_1^T.
\end{eqnarray*}
Now, $U_1^T\tX_\varepsilon^T\tX_\varepsilon U_1\succ0$ since
$U_1^T\tX_\varepsilon^T\tX_\varepsilon U_1=U_1^T\tX^T\tX U_1+J$ and
$J\succeq0$
and $\tX U_1$ has full column rank. Thus,
$\ns(I_{1:n}(P\tX_\varepsilon^T)^+)=\ns(U_1^T)=\ns(\tX P)$. This
completes the first part of the proof.

Next, we show that $\tX P(e_j^+-e_j^{-})\neq0$ as long as $[(\tX_\varepsilon
P)^+(y_\varepsilon-\break(P\tX_\varepsilon^T)^+w)^T]\times e_j^\pm>0$.
Now, since $j\in\cap\P(\bp)\cap\P(\bn)\subseteq\T$, the only row of $\tD$ that has $\tD_i(e_j^\pm)\neq0$ is the R1
row; but clearly $\tD_i(e_j^+-e_j^-)=1-1=0$. Thus,
$e_j^+-e_j^-\in\ns(\tD)$ and $P(e_j^+-e_j^-)=e_j^+-e_j^-$. It
follows that
\[
\tX P\bigl(e_j^+-e_j^-\bigr)= \pmatrix{
2x_j\vspace*{2pt}
\cr
\sqrt{\varepsilon}e_j
\vspace*{2pt}\cr
-\sqrt{\varepsilon }e_j\vspace *{2pt}
\cr
0 }\neq0 \qquad\mbox{assuming }
\varepsilon>0.
\]

Putting these two parts of the proof together establishes that\break
$I_{1:n}(P\tX_\varepsilon^T)^+(e_j^+-e_j^-)\neq0$. Thus,
$\{y_\varepsilon\in\N\}$ is a finite union of Lebesgue measure 0 sets.
This shows that $P(y_\varepsilon\in\N)=0$ as long as $y$ is absolutely
continuous with respect to the Lebesgue measure on $\real^n$.

\subsection{\texorpdfstring{Proof of weak hierarchy (Property \protect\ref{prop:weak-hier})}
{Proof of weak hierarchy (Property 3)}}
\label{sec:prov-weak-hier}

An argument nearly identical to that of the previous section
establishes that
$\hTheta_{jk}\neq0\implies\hbp_j-\hbn_j\neq0$ with probability one.
Thus, if $\hbp_j-\hbn_j=0$ and $\hbp_k-\hbn_j=0$, then both
$\hTheta_{jk}=0$ and $\hTheta_{kj}=0$. It follows then that
$(\hTheta_{jk}+\hTheta_{kj})/2=0$. This establishes weak hierarchy.

\section{Degrees of freedom}
\label{app:df}

\subsection{\texorpdfstring{Proof of unbiased estimate (Property \protect\ref{prop:df})}
{Proof of unbiased estimate (Property 4)}}
\label{sec:proof-unbi-estim}

The fit in terms of the active set is given by
\[
\tX\hphi=(\tX P) (\tX P)^+\bigl(y-\bigl(P\tX^T\bigr)^+w\bigr),
\]
where $P=P_{\ns(L)\cap\ns(D_{-\A(\hphi)})}$.
Of course, $\hmu_{\A(\hphi)}=0$, and we can solve the KKT conditions to
get the rest of the optimal dual
variables in terms of the active set
\[
\pmatrix{ \hmu_{-\A(\hphi)}\vspace*{2pt}
\cr
\hnu } = \tD^{T+}\bigl[w-
\tX^T(y-\tX\hphi)\bigr]+c,
\]
where $c\in\ns(\tD)$ satisfies $\tD^{T+}[w-\tX^T(y-\tX\hphi
)]+c\ge0.$

Note that $\hphi=\hphi(y)$ and thus $\A(\hphi)$ and $b$ depend on
$y$ even
though we do not write this explicitly. We will continue writing
$\hphi$ to mean specifically $\hphi(y)$. For $y'$ in a neighborhood of
$y$, we might guess that $\hphi(y')=f(y')$ and $(\hmu(y'),\hnu
(y'))=(g(y'),h(y'))$, where
\begin{eqnarray*}
f\bigl(y'\bigr)&=&(\tX P_{\ns(L)\cap\ns(D_{-\A(\hphi)})})^+\bigl(y'-
\bigl(P_{\ns(L)\cap\ns
(D_{-\A(\hphi
)})}\tX^T\bigr)^+w\bigr)\\
&&{}+b,
\\
\pmatrix{ g\bigl(y'\bigr)_{-\A(\hphi)}\vspace*{2pt}
\cr
h
\bigl(y'\bigr) } &=& \pmatrix{ D_{-\A(\hphi)}\vspace*{2pt}
\cr
L
}^{T+}\bigl[w-\tX^T\bigl(y'-\tX f
\bigl(y'\bigr)\bigr)\bigr]+c,
\\
g\bigl(y'\bigr)_{\A(\hphi)} &=& 0.
\end{eqnarray*}
To verify this guess, we need to check that the pair
$(f(y'),(g(y'),h(y')))$ satisfies the optimality conditions
at $y'$,
\begin{eqnarray*}
\tX^T\bigl(y'-\tX f\bigl(y'\bigr)
\bigr)&=&w-D^T g\bigl(y'\bigr)+L^T h
\bigl(y'\bigr),
\\
g_i\bigl(y'\bigr) \bigl(D f\bigl(y'\bigr)
\bigr)_i&=&0,
\\
g\bigl(y'\bigr)&\ge&0,\qquad Df\bigl(y'\bigr)\ge0,\qquad Lf
\bigl(y'\bigr)=0.
\end{eqnarray*}
First of all, $Lf(y')=0$ holds since $L(\tX
P_{\ns(L)\cap\ns(D_{-\A(\hphi)})})^+=0$ and $Lb=0$. Likewise,
$D_{-\A(\hphi)}f(y')=0$. Now $D_{\A(\hphi)}f(y)>0$, so by
continuity of
$f$, we have $D_{\A(\hphi)}f(y')>0$ for all $y'$ in a
small enough neighborhood, $U_1$, of $y$. This establishes
that $\A(f(y'))=\A(\hphi)$. Now $g(y')_{\A(\hphi)}=0$, so complementary
slackness holds. To see that the first optimality condition holds, we
can simply plug $(g(y'),h(y'))$ into the left-hand side. All that
remains is to show that $g(y')_{-\A(\hphi)}\ge0$. If we knew that
$\hmu_{-\A(\hphi)}>0$, then by continuity of $g$ we could argue that
over a small enough neighborhood, $U_2$, $g(y')_{-\A(\hphi)}>0$.
However, it could be the case that $\hmu_i=0$ for some
$i\notin\A(\hphi)$, that is, $i\in\B(\hphi)\setminus\A(\hphi)$.
Nonetheless, one can show that there is a set $\cal N$ of measure 0 for which
$y\notin\cal N$
implies that $\A(\hphi(y))=\A(\hphi(y'))$ and $\B(\hphi(y))=\B
(\hphi
(y'))$ for all $y'$ in a neighborhood of $y$. Lemma 9 of (Ryan)
\citet{RyanTibs12} proves this result for a nearly identical
situation.

The fit $\tX\hphi(y)$ is a piecewise affine function of $y$. Using
Stein's formula for the degrees of freedom [as described in Ryan,
\citet{RyanTibs12}], we get that
\[
df(\tX\hat\phi)=E\bigl[\nabla\cdot\tX\hphi(y)\bigr]=E\bigl[\tr\bigl\{(\tX P) (
\tX P)^+\bigr\} \bigr]=E\bigl[\operatorname{rank}(\tX P)\bigr],
\]
where
$P=P_{\ns(L)\cap\ns(D_{-\A(\hphi)})}$.

\subsection{\texorpdfstring{Proof of bound on estimate (Property \protect\ref{prop:df-bound})}
{Proof of bound on estimate (Property 5)}}
\label{sec:proof-bound-estimate}

We bound this by an estimate that is more interpretable: $\operatorname
{rank}(\tX P)\le
\operatorname{rank}(P)=\mathrm{nullity}{
L\choose D_{-\A(\hphi)}}$.

Clearly, R2--R7 are linearly independent rows. Thus, the rank of
$\widetilde D$ is at least
$|\Z(\bp)|+|\Z(\bn)|+|\Z(\Theta^+)|+|\Z(\Theta^-)|+2p$. Now, an R1
row is linearly independent of R2--R8 precisely when $j\in\T$
has $j\in\Z(\bp)\Delta\Z(\bn)$. To see this, note that if
$j\in\Z(\bp)\setminus\Z(\bn)$, then R1 is certainly linearly
independent of R2--R8 and likewise for $j\in\Z(\bn)\setminus\Z(\bp
)$; however
if $j\in\Z(\bp)\cap\Z(\bn)$, then $jk\in\Z(\Theta^+)\cap\Z
(\Theta^-)$ for
all $k\in\{1,\ldots,p\}\setminus\{j\}$, and therefore this row of R1
lies in the span of R3--R7. Thus, this means there are
$|\T\setminus(\Z(\bp)\Delta\Z(\bn))|$ additional linearly independent
rows. Finally, we consider R8. Clearly, R8 lies in the span of
R4--R5 for $jk\in\Z(\Theta^+)\cap\Z(\Theta^-)$ since
$jk\in\Z(\Theta^+)\implies kj\in\Z(\Theta^+)$ at a solution. But if
$jk\in\P(\Theta^+)\cup\P(\Theta^-)$, then it is linearly independent
of R1--R8. Therefore, R8 adds $|\P(\Theta^+)|/2+|\P(\Theta^-)|/2$ to
the rank where we have used that $\P(\Theta^+)\cap\P(\Theta
^-)=\varnothing
$ at
a solution (since $\lambda_2>0$) and recalling that $j < k$ for the
rows of R8. In summary, we have shown that the row-rank is
\begin{eqnarray*}
&&\bigl|\Z(\bp)\bigr|+\bigl|\Z(\bn)\bigr|+\bigl|\Z\bigl(\Theta^+\bigr)\bigr|+\bigl|\Z\bigl(\Theta^-\bigr)\bigr|+2p
\\
&&\qquad{}+ \bigl|\T\setminus\bigl(\Z(\bp)\Delta\Z(\bn)\bigr)\bigr|+\bigl|\P\bigl(\Theta^+\bigr)\bigr|/2+\bigl|\P
\bigl(\Theta^-\bigr)\bigr|/2.
\end{eqnarray*}
Recalling that there are $2p+2p^2$ columns, we get that
\begin{eqnarray*}
&&\mathrm{nullity}\pmatrix{ L\vspace*{2pt}
\cr
D_{-\A(\hphi)} }=
\bigl(2p+2p^2\bigr)-\operatorname{rank}\pmatrix{ L\vspace*{2pt}
\cr
D_{-\A(\hphi)} }
\\
&&\qquad=\bigl(p-\bigl|\Z(\bp)\bigr|\bigr)+\bigl(p-\bigl|\Z(\bn)\bigr|\bigr)+\bigl(p(p-1)-\bigl|\Z\bigl(\Theta^+
\bigr)\bigr|\bigr)\\
&&\qquad\quad{}+\bigl(p(p-1)-\bigl|\Z \bigl(\Theta ^-\bigr)\bigr|\bigr)
\\
&&\qquad\quad{} -\bigl|\T\setminus\bigl(\Z(\bp)\Delta\Z(\bn)\bigr)\bigr|-\bigl|\P\bigl(\Theta ^+\bigr)\bigr|/2-\bigl|\P
\bigl(\Theta ^-\bigr)\bigr|/2
\\
&&\qquad=\bigl|\P(\bp)\bigr|+\bigl|\P(\bn)\bigr|+\bigl|\P\bigl(\Theta^+\bigr)\bigr|/2+\bigl|\P\bigl(\Theta^-\bigr)\bigr|/2\\
&&\qquad\quad{}-\bigl|\T
\setminus\bigl(\P (\bp)\Delta\P(\bn)\bigr)\bigr|.
\end{eqnarray*}

\section{Solving the prox function}
\label{app:prox}
The Lagrangian of \eqref{eq:prox} is given by
\begin{eqnarray*}
L\bigl(\beta^\pm,\theta;\alpha,\gamma^\pm\bigr)&=&
\frac{1}{2t}\bigl(\bp-\tilde \beta ^+\bigr)^2+\frac{1}{2t}
\bigl(\bn-\tilde\beta^-\bigr)^2+\frac{1}{2t}\Vert\theta -\tilde
\theta\Vert^2
\\
&&{}+ \biggl(\frac{\lambda}{2}+\alpha \biggr)\Vert\theta\Vert _1-\bigl(\gp
+\alpha\bigr)\bp-\bigl(\gn+\alpha\bigr)\bn,
\end{eqnarray*}
where $\alpha$ is the dual variable corresponding to the hierarchy
constraints and $\gamma^\pm$ are the dual variables corresponding to
the nonnegativity constraints. For notational convenience, we have written
$\theta$ for $\Theta_j$, and we have dropped the subscripts on
$\beta^\pm$. The KKT conditions are
\begin{eqnarray*}
\bigl(\hat\beta^\pm-\tilde\beta^\pm\bigr)/t-\hat
\gamma^\pm-\hat\alpha &=&0,\qquad (\hat\theta-\tilde\theta)/t+(\lambda/2+\hat
\alpha)u=0,
\\
\hat\beta^\pm\hat\gamma^\pm&=&0,\qquad \hat\alpha\bigl(\Vert\hat\theta
\Vert_1-\hbp-\hbn\bigr)=0,
\\
\hat\beta^\pm&\ge&0, \qquad\Vert\hat\theta\Vert_1\le\hbp+\hbn,\qquad
\hat\alpha\ge0,\qquad \hat\gamma^\pm\ge0,
\end{eqnarray*}
where $u_j$ is a subgradient of the absolute value function evaluated
at $\hat\theta_j$. The three conditions involving $\gamma^\pm$
implies that $\hat\beta^\pm=[\tilde\beta^\pm+t\hat\alpha]_+$. The
stationarity condition involving $\hat\theta$ implies that
$\hat\theta=S(\tilde\theta, t(\lambda/2+\hat\alpha))$. Now, define
$f(\alpha)=\Vert S(\tilde\theta, t(\lambda/2+\alpha))\Vert
_1-[\tilde
\beta^++t\alpha]_+-[\tilde\beta^-+t\alpha]_+.$
The remaining KKT conditions involve $\hat\alpha$ alone: $\hat\alpha
f(\hat\alpha) = 0, f(\hat\alpha)\le0, \hat\alpha\ge0.$
Observing that $f$ is nonincreasing in $\alpha$ and piecewise
linear suggests finding
$\hat\alpha$ as done in Algorithm~\ref{alg:onerow}.

\begin{algorithm}[t]
\caption{\texttt{ONEROW}: Solve \protect\eqref{eq:prox} via dual.}\label{alg:onerow}
\emph{Inputs:} $\tilde\beta^+_j,\tilde\beta^-_j\in\real, \widetilde
\Theta_j\in\real^{p-1}, \lambda\ge0$.
\begin{enumerate}
\item[(1)] Find $\hat\alpha.$
Define $f(\alpha)=\Vert S(\widetilde\Theta_j, t(\lambda/2+\alpha
))\Vert
_1-[\tilde\beta_j^++t\alpha]_+-[\tilde\beta_j^-+t\alpha]_+.$
\begin{enumerate}
\item[(a)] If $f(0)\le0$, take $\hat\alpha=0$ and go to step 2.
\item[(b)] Form knot the set $\mathcal
P=\{|\widetilde\Theta_{jk}|/t-\lambda/2\}_{k=1}^p\cup\{-\tilde
\beta^\pm
/t\},$
and let $\mathcal P^+=\mathcal P\cap[0,\infty)$.
\item[(c)] Evaluate $f(p)$ for $p\in\mathcal P^+$.
\item[(d)] If $f(p) = 0$ for some $p\in\mathcal P^+$, take $\hat\alpha=p$
and go to step 2.
\item[(e)] Find adjacent knots, $p_1,p_2\in\mathcal P^+$, such that
$f(p_1)>0>f(p_2)$. Take
\[
\hat\alpha=-f(p_1)\bigl[f(p_2)-f(p_1)
\bigr]/(p_2-p_1).
\]
\end{enumerate}
\item[(2)] Return $\hTheta_j=\mathcal
S[\widetilde\Theta_j, t(\lambda/2+\hat\alpha)]$ and $\hat\beta
^\pm
_j=[\tilde\beta^\pm_j+t\hat\alpha]_+$.
\end{enumerate}
\end{algorithm}

\end{appendix}

\section*{Acknowledgments}
We thank Will Fithian, Max Grazier-G'Sell, Brad Klingenberg,
Balasubramanian Narasimhan and Ryan
Tibshirani for useful conversations and two referees and an Associate
Editor for helpful comments.

\begin{supplement}[id=suppA]
\stitle{Supplement to ``A lasso for hierarchical interactions''\\}
\slink[doi]{10.1214/13-AOS1096SUPP} 
\sdatatype{.pdf}
\sfilename{aos1096\_supp.pdf}
\sdescription{We include proofs of Property~\ref{prop:constraint}
and of the statement in Remark~\ref{re3}. Additionally, we show that the
algorithm for the logistic regression case is nearly identical and
give more detail on Algorithm~\ref{alg:admm}.}
\end{supplement}

%

\printaddresses

\end{document}